\documentclass[useAMS,usenatbib,usegraphicx]{mn2e} 
\usepackage[usenames]{color}
\usepackage{amssymb}
\usepackage{amsmath}
\usepackage{bm}

\newcommand{\be}{\begin{equation}}
\newcommand{\ee}{\end{equation}}
\newcommand{\bea}{\begin{eqnarray}}
\newcommand{\eea}{\end{eqnarray}}
\newcommand{\lm}{\Lambda{\rm CDM}}

\newcommand{\cmass}{\,M_\odot\,h^{-1}}
\newcommand{\hdist}{\,h^{-1}{\rm Mpc}}
\newcommand{\kms}{\,{\rm km\,s}^{-1}}
\newcommand{\MR}{M|\lambda}

\title[redMaPPer vs MXXL]{Precise Clustering and Density Evolution of
redMaPPer Galaxy Clusters versus MXXL Simulation}
\author[Jimeno et al.]
{Pablo Jimeno$^{1}$\thanks{E-mail: pablodavid.jimeno@ehu.eus}, Tom Broadhurst$^{1,2}$, Ruth Lazkoz$^{1}$, Raul Angulo$^{3}$,
\newauthor
Jose-Maria Diego$^{4}$, Keiichi Umetsu$^{5}$, Ming-chung Chu$^{6}$\\
$^{1}$Department of Theoretical Physics and History of Science, University of the Basque Country UPV-EHU, 48040 Bilbao, Spain\\
$^{2}$IKERBASQUE, Basque Foundation for Science, Alameda Urquijo, 36-5 48008 Bilbao, Spain\\
$^{3}$Centro de Estudios de F\'isica del Cosmos de Arag\'on (CEFCA), Plaza San Juan 1, Planta-2, 44001, Teruel, Spain\\
$^{4}$IFCA, Instituto de F\'isica de Cantabria (UC-CSIC), Av. de Los Castros s/n, 39005 Santander, Spain\\
$^{5}$Institute of Astronomy and Astrophysics, Academia Sinica, P.O. Box
23-141, Taipei 10617, Taiwan\\
$^{6}$ Department of Physics, The Chinese University of Hong Kong, Shatin, N.T., Hong Kong, China\\
}

\date{\today} 

\pagerange{\pageref{firstpage}--\pageref{lastpage}}

\begin{document}

\maketitle

\label{firstpage} 

%%%%%%%%%%%%%%%%%%%%%%%%%%%%%%%%%%%%%%%%%%%%%%%%%%%%%%%%%%%%%%%%%%%%%%%%%%%%%%%
\begin{abstract} 
We construct a large, redshift complete sample of distant galaxy
 clusters by correlating Sloan Digital Sky Survey (SDSS) Data Release 12
 (DR12) redshifts with clusters identified with the red-sequence Matched-filter
 Probabilistic Percolation (redMaPPer) algorithm. 
Our spectroscopic completeness is $>97\%$ for $\simeq 7000$ clusters
 within the redMaPPer selection limit, $z\leqslant0.325$, so that our
 cluster correlation functions are much more precise than earlier work
 and not suppressed by uncertain photometric redshifts. We derive an accurate
 power-law mass--richness relation from the observed abundance with
 respect to the mass function from Millennium XXL (MXXL) simulations,
 adjusted to the {\em Planck} weighted cosmology. The number density of
 clusters is found to decline by 20\% over the range $0.1<z<0.3$, in
 good agreement with the evolution predicted by MXXL. Our projected
 three-dimensional correlation function scales with richness, $\lambda$, rising from
 $r_0=14\hdist$ at $\lambda\simeq 25$, to $r_0=22\hdist$ at
 $\lambda\simeq 60$, with a gradient that matches MXXL when applying our
 mass--richness relation, whereas the observed amplitude of the
 correlation function at $\left<z\right>=0.24$ exceeds the MXXL
 prediction by 20\% at the $\simeq 2.5\sigma$ level. This tension cannot
 be blamed on spurious, randomly located clusters as this would reduce
 the correlation amplitude. Full consistency between the correlation
 function and the abundances is achievable for the pre-{\em Planck}
 values of $\sigma_8=0.9$, $\Omega_m=0.25$, and $h=0.73$, matching the
 improved distance ladder estimate of the {\em Hubble} constant.
\end{abstract}
%%%%%%%%%%%%%%%%%%%%%%%%%%%%%%%%%%%%%%%%%%%%%%%%%%%%%%%%%%%%%%%%%%%%%%%%%%%%%%%

\begin{keywords} 
cosmology: observations ---
dark matter ---
galaxies: clusters: general ---
gravitational lensing: weak ---
large-scale structure of Universe
\end{keywords} 
%%%%%%%%%%%%%%%%%%%%%%%%%%%%%%%%%%%%%%%%%%%%%%%%%%%%%%%%%%%%%%%%%%%%%%%%%%%%%%%

%%%%%%%%%%%%%%%%%%%%%%%%%%%%%%%%%%%%%%%%%%%%%%%%%%%%%%%%%%%%%%%%%%%%%%%%%%%%%%%
%%%%%%%%%%%%%%%%%%%%%%%%%%%%%%%%%%%%%%%%%%%%%%%%%%%%%%%%%%%%%%%%%%%%%%%%%%%%%%%
\section{Introduction}
\label{sect:introduction} 
%%%%%%%%%%%%%%%%%%%%%%%%%%%%%%%%%%%%%%%%%%%%%%%%%%%%%%%%%%%%%%%%%%%%%%%%%%%%%%%

The celebrated sensitivity of cluster abundance to the growth rate of structure means that the mass density of the Universe, $\Omega_m$, and the amplitude of the power spectrum, $\sigma_8$ should be particularly accurately derived from cluster surveys, \citep{Peebles1989, Oukbir1992, Eke1996, Viana1996, Oukbir1997, Carlberg1997, Bahcall1997, Fan1997, Henry1997, Voit2005, Brodwin2007, Mantz2010, Rozo2010, Clerc2012, Benson2013, Huterer2015} providing a welcomed consistency check of the current cosmology. Even the presence of a few massive clusters at $z>0.5$ has been enough to overcome the long desired $\Omega_m=1$ consensus \citep{Bahcall1997}, favouring a sub-critical mean mass density for the Universe \citep{Bahcall1998}. Empirically, $\Omega_m\simeq 0.2$--$0.3$ has long been argued by simply extrapolating galaxy mass-to-light ratios (M/L) to large scales \citep{Ostriker1974} and clarified with dynamical measurements on larger scales \citep{Bahcall1995, Peacock2001} and of course independently confirmed with increasingly accurate claims from CMB acoustics \citep{Spergel2003, Planck2015XIII}.

 New underway surveys to find large samples of clusters above $z \geqslant 0.5$ with lensing based masses are very exciting in this respect, so growth can be tracked as a function of cluster mass with unprecedented precision, including subtle modification by cosmological neutrinos \citep{Younger2005, Lahav2010, Carbone2012, Weinberg2013, Leistedt2014}, with little complication anticipated from detailed cosmological simulations that include gas physics \citep{Bocquet2016}. Using clusters, the current best estimates of the $\sigma_8\,\Omega_m^{\simeq 0.5}$ combination that principally determines growth \citep{Kaiser1984, Rozo2010, Mandelbaum2013} has hitherto been limited to the local volumes where cosmic variance and relatively small samples means it is rather unclear how to assess differences with the CMB based $\Omega_m h^2$ combination, fixed principally by the first peak of the CMB, and $\sigma_8$, where the uncertain level of electron scattering, $\tau$, smooths the amplitude of CMB fluctuations. The {\em Planck} weighted values of these observationally interdependent parameters, $\sigma_8$, $\Omega_m$, $h$, $\tau$, are now claimed to be in significant tension with the constantly high value of $H_0$ derived locally from the distance ladder \citep{Riess2016}.

Undermining the use of clusters in such comparisons is the indirectness of cluster mass estimates for which empirical scaling relations have to be relied on for converting observables to mass. Cluster richness seems to provide a robust connection as it is close to being linearly related to mass \citep{Rozo2009a, Rozo2009b, Rykoff2012}, with a slope of $d\log M_{200}/d\log N \simeq 1.1$, and a $\simeq 20\%$ inherent scatter inferred \citep{Andreon2016b} and with little evidence of evolution \citep{Younger2005, Andreon2014, Hennig2016}. Power-law scalings to convert X-ray and SZ measurements to total cluster masses are complicated by compressed gas and shocks from cluster interactions, so that the selection function and its evolution is challenging \citep{Suto2000, Allen2011, Mroczkowski2012, Molnar2015, Sereno2015a, Andreon2016a}. Lensing based scaling relations are now feasible for limited samples but for which the initial X-ray or SZ selection complicates matters \citep{Rozo2014b, Sereno2015c}.

Deeper, higher resolution, wide-field imaging surveys that exceed the SDSS, should soon rectify this lack of clusters with masses based on lensing and finally provide the long hoped statistically large sample of ``mass selected'' clusters to greater depth, in particular the Subaru/Hyper Suprime-Cam (HSC) \citep{Takada2010, Miyazaki2015}, J-PAS Northern Sky \citep{Benitez2014} and DES \citep{DES2016} surveys now underway. Broad-band surveys like HSC will require careful avoidance of foreground/member dilution of the background lensing signal that is feasible by excluding degenerate colour space \citep{Broadhurst2005, Medezinski2007, Medezinski2010, Umetsu2008}, with the fullest wavelength coverage to maximise the numbers of galaxies redder than the cluster and also dropouts. In the case of J-PAS the many narrow bands will provide unambiguous redshifts based on resolved spectral features, allowing clusters to be identified cleanly and to relatively low mass \citep{Ascaso2016}. Such data will improve future and past SZ, X-ray and radio ``relic'' based cluster searches \citep{Rottgering2011}, which will benefit enormously from lensing masses and the definition of cluster membership \citep{Hennig2016}, and can enhance the utility of the SDSS in the North.

Beyond cluster abundances, higher order moments of the density field, including the correlation function of galaxy clusters, also relate directly to the growth of structure \citep{Brodwin2007}. The clustering of clusters is in this respect far more useful than for galaxies where the strong dependence on Hubble-type implies a complex ``astrophysical bias'' \citep{Peacock2001, Lahav2002}. A major advantage of using clusters is their clear relation to the mass distribution, especially if direct lensing masses can be obtained for statistically large samples of clusters. This is unlike galaxies where ellipticals are measured to be much more spatially correlated than disk galaxies, implying as may be expected that the creation of galaxies from the underlying mass distribution is not simply related to the local density of dark matter. In the case of clusters the bias is more simply mass-density related and is not expected to be significantly influenced by gas physics, allowing relatively clean comparisons between theory and observation. For clusters a nearly linear relation is established between the measured richness and mass with a modest scatter, so richness can be reliably transformed statistically when examining the clustering of clusters. Previous clustering work with the SDSS has been either with relatively small local samples with an uncertain mass--richness relation, or relies on corrections for the wide smoothing by photometric redshifts \citep{Sereno2015c}, or on the angular clustering \citep{Baxter2016}. Here we establish the first spectroscopically complete analysis of cluster statistics using the depth of the SDSS survey, beyond the local Universe.

The careful redMaPPer work has been a big advance in identifying clusters by their red sequence of member galaxies and deriving reliable richnesses using the SDSS/BOSS survey data which has sufficient depth to detect clusters to $z~\simeq 0.3$ with high completeness \citep{Rykoff2014}. Currently 70\% of the brightest cluster galaxies in this clusters have redshift measurements with the SDSS/DR12 release. Here we augment these BCG measurements with additional cluster member redshift measurements, by correlating redMaPPer identified red sequence galaxies with the enlarged DR12 redshift sample from SDSS/BOSS, which we show here provides spectroscopic completeness to 93\% overall, and $>97$\% for the redMaPPer richness complete redshift range $z<0.325$ that we focus on in this paper.

In tandem with this observational progress, advances in the N-body simulations of $\lm$ have extended to volumes several times that of the observable Universe \citep{Angulo2012}. Large simulated volumes are necessary to accurately predict the number of massive clusters, given their rarity. The cosmological parameters chosen for these simulations follow the tradition set by \cite{Springel2005a} for such ground breaking simulations allowing consistency checks between these generations of simulations. The former consensus values adopted for these simulations \citep{Seljak2005} differ significantly from the present {\em Planck} weighted values of $\sigma_8$ and $\Omega_m$ that principally influence cluster predictions. The cause of this may be traced mainly to the relatively large $\tau$ estimated by WMAP \citep{Spergel2003}, raising $\sigma_8=0.9$, and lower $\Omega_m=0.25$. The amplitude of the CMB fluctuations on large scales scale as $A_s\,exp(-2\tau)$ where $A_s$ is the amplitude of the matter power spectrum; hence a higher $\tau$ implies a higher $A_s$ and consequently a higher $\sigma_8 \propto A_s$. The {\em Planck} weighted values today are significantly ``reversed'' for these key parameters mainly because of the much lower inferred $\tau$. 

The structure of this paper is as follows. In Sect.~\ref{sect:data} we describe the data, namely, the redMaPPer cluster catalogue and the SDSS/BOSS spectroscopic sample. In Sec.~\ref{sect:mxxl} we describe the MXXL simulations that we use for comparison, and in Sec.~\ref{sect:correlation} and Sec.~\ref{sect:abundances} we describe the results obtained from the clustering and abundances analyses, respectively. Then, in Sec.~\ref{sect:likelihood} we perform a likelihood analysis to obtain the cosmologically favoured mass--richness relation, and explore some dependence on cosmology and its consistency with the clustering and the abundances results. In Sec.~\ref{sect:redshift_enhancement} we present the latest measurements of the redshift enhancement effect. Finally, we present our conclusions in Sec.~\ref{sect:conclusions}. Throughout we compare both the \cite{Planck2015XIII} weighted cosmological parameters and the consensus parameters set in 2003 used for the largest available simulations that differ significantly in terms of $\sigma_8$, $\Omega_m$ and $H_0$.

%%%%%%%%%%%%%%%%%%%%%%%%%%%%%%%%%%%%%%%%%%%%%%%%%%%%%%%%%%%%%%%%%%%%%%%%%%%%%%%
%%%%%%%%%%%%%%%%%%%%%%%%%%%%%%%%%%%%%%%%%%%%%%%%%%%%%%%%%%%%%%%%%%%%%%%%%%%%%%%
\section{Data}
\label{sect:data}
%%%%%%%%%%%%%%%%%%%%%%%%%%%%%%%%%%%%%%%%%%%%%%%%%%%%%%%%%%%%%%%%%%%%%%%%%%%%%%%

All our observational data comes from the Sloan Digital Sky Survey (SDSS), the most successful photometric and spectroscopic survey conducted to date \citep{Gunn2006}. Since it began in 2000, the SDSS has mapped the largest portion of the Universe to date, and provides high-precision data that has proved very useful for several kinds of large-scale structure analyses. The telescope has scanned more than 14,000 deg$^2$ of the sky with a mosaic CCD with five color-bands, {\it u, g, r, i } and {\it z} \citep{Fukugita1996}. It has also obtained the spectra of more than 1,600,000 unique objects during the SDSS-I/II progammes \citep{York2000}, and more than 1,500,000 unique spectra with a more advanced 1,000-fiber spectrograph \citep{Smee2013} during SDSS-III \citep{Eisenstein2011}, completed on 2014. The information obtained in this survey has been made public to the scientific community on a series of different Data Releases (DR), being the latest DR12 \citep{Alam2015}, which contains, among other information, 14,500 deg$^2$ of imaging that includes photometric information of 208,478,448 galaxies, and the optical spectra of 2,401,952 unique galaxies to $z=0.7$ in $\sim$10,500 deg$^2$ of the sky. 

\subsection{redMaPPer Cluster Catalogue}

The red-sequence Matched-filter Probabilistic Percolation (redMaPPer, \citealp{Rykoff2014}) is the most robust and complete cluster catalogue based on SDSS data that has been produced to date. The redMaPPer cluster finder algorithm relies on a self-training procedure that calibrates the red-sequence as a function of redshift from a sample of galaxy clusters with known red spectroscopic galaxies. This red-sequence pattern is then used in the photometric data to find potential clusters and find, through an iterative process, the central galaxy (CG), the redshift $z_\lambda$, and the richness $\lambda$ of each cluster. This richness estimator is a multi-color evolution of the previous estimators described in \cite{Rozo2009b} and \cite{Rykoff2012}, which where employed in maxBCG \citep{Koester2007a} and GMBCG \citep{Hao2010} catalogues.

The richness of a cluster is defined as:
\begin{equation}
\lambda=\sum p_i\,\theta_i^L\,\theta_i^R\,,
\end{equation}
where $p_i$ is the probability that each galaxy found near the cluster is actually a cluster member, and $\theta_i^L$ and $\theta_i^R$ are the luminosity and radius-dependant optimized weights:
\begin{equation}
\theta_i^L=\frac{1}{2}\left[1+{\rm erf} \left( \frac{m_{\rm max}-m_i}{\sigma_i}\right) \right]\,,
\end{equation}
\begin{equation}
\theta_i^R=\frac{1}{2}\left[1+{\rm erf} \left( \frac{R(\lambda)-R}{\sigma_R}\right) \right]\,,
\end{equation}
with $m_{\rm max}$ the magnitude that corresponds to the $0.2 L_*$ luminosity threshold, $\sigma_i$ the photometric error of galaxy $i$, $\sigma_R=0.05 \hdist$, and $R(\lambda)$ the richness-dependant aperture:
\begin{equation}
R(\lambda)=\left( \frac{\lambda}{100}\right)^{0.2}\hdist\,.
\end{equation}

We refer the reader to \cite{Rykoff2014} and \cite{Rozo2015b} for an in-depth explanation of the algorithm features.

This cluster finder algorithm was designed to process future large photometric surveys like DES and Large Synoptic Survey Telescope (LSST), but it has been already run on the SDSS DR8 photometric and SDSS DR9 spectroscopic data in order to check the possible systematics associated to the algorithm itself, and compare its results with other clusters catalogues, like the SZ {\em Planck} cluster catalogue \citep{Rozo2015a}, or the X-CLASS X-ray cluster \citep{Sadibekova2014}. These analyses led to an updated 5.10 version of the algorithm described in \cite{Rozo2015b}, which we will augment here with the latest DR12 release that we show below provides a substantial enhancement in the number of the clusters with reliable spectroscopic redshifts.

The resulting redMaPPer catalogue covers an effective area of $\sim$10,400 deg$^2$, and contains 26,350 clusters in the $0.08 \leqslant z_{\rm photo} \leqslant 0.55$ redshift range.

There are several reasons why we select this catalogue among all the others described in the literature:

\begin{itemize}

 \item High purity, where purity here is not understood as false detections (like is usually defined in SZ and X-ray cluster samples), but as the fraction of clusters that is not affected by projection effects that may lead to overestimated richness measurements. This purity is claimed to be $>95\%$ in \cite{Rykoff2014}, although a more recent study \citep{Simet2016} finds a higher rate of projection effects of the order of $12\%\pm4\%$.
 
 \item A conservative low richness cutoff, or detection threshold of $\lambda/S(z)>20$, that enhances the performance of the resulting cluster catalogue. The ``scale factor'' $S(z)$ is introduced in order to take into account the limited depth of the sample, so a cluster of richness $\lambda$ has $\lambda/S(z)$ galaxies above the magnitude limit of the survey, and $S=1$ at $z<0.35$, where the DR8 is volume limited. This richness cut corresponds to a mass limit of approximately $M_{200c}\geqslant1.4\times10^{14}\cmass$.
 
 \item Together with the photometric redshift estimates of each cluster, an SDSS spectroscopic redshift measurement is provided for the CG when available. Of the original 26,350 clusters contained in the catalogue, 16,259 contain this information.
 
 \item Instead of providing a unique CG candidate for the cluster, the algorithm indicates the centering probability $p_{\rm cen}$ of the five most probable CGs, together with their position in the cluster.
 
 \item As previous studies \citep{Jimeno2015} that checked and compared different cluster catalogues have shown, redMaPPer is the catalogue that offered the most consistent results in terms of a cross comparison of the redshift enhancement and the gravitational redshift effects associated to clusters.
 
 \item Provides supplementary information of the member galaxies that have been considered in the richness estimation, with their membership probability. There is information for 1,736,221 member galaxies, of which 72,642 also contain spectroscopic redshift measurements.
 
\end{itemize}

\begin{figure}
\resizebox{84mm}{!}{\includegraphics{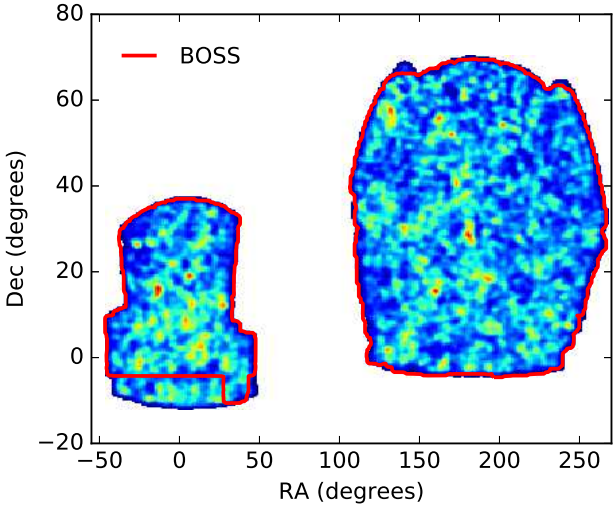}}
\caption{redMaPPer catalogue and BOSS footprint. The region inside the red line represents the area of the sky covered by the BOSS that overlaps with the redMaPPer catalogue. As can be seen from the figure, there is a region in the bottom left corner with redMaPPer clusters outside the BOSS region. We exclude those clusters from the final sample so that we have uniformly high completeness between the redMaPPer and the BOSS samples required in our analysis.}
\label{fig:redMaPPer_Sky_Distribution}
\end{figure}

\subsection{BOSS spectroscopic sample}

The Baryon Oscillations Spectroscopic Survey (BOSS, \citealp{Dawson2013}), designed to measure the baryon acoustic oscillations (BAO) scale at $z=0.3$ and $z=0.57$ to a 1.0\% accuracy, is the largest of the four surveys that comprise the SDSS-III progamme. Now completed, it uniformly targeted and obtained the spectra of galaxies in two redshift ranges: $0.15<z<0.4$, which lead to the color-selected ``LOWZ'' sample, composed of the brightest and the reddest of the low redshift galaxies, and $0.4<z<0.8$, designed to obtain through a series of photometric colour cuts a volume-limited sample of galaxies with approximately constant stellar mass, the so called ``CMASS'' sample. A total of about a half and a million unique galaxy spectra were measured over 10,400 deg$^2$ of the sky in the Northern and Southern Galactic Caps.

We update the publicly available catalogue based on DR8 and DR9 data using the more recent DR12 BOSS spectroscopic data. In order to work with the most reliable data, from all the BOSS galaxy spectra we select those that satisfy the following quality flag conditions: \texttt{PLATEQUALITY}$\,=\,$\texttt{good}, \texttt{Z\_ERR\_NOQSO}$\,<0.001$ and \texttt{ZWARNING\_NOQSO}$\,=0$ or \texttt{ZWARNING\_NOQSO}$\,=16$. The footprint of the final considered sample, which contains the spectra of 1,339,107 galaxies, is shown in Fig.~\ref{fig:redMaPPer_Sky_Distribution}. We cross correlate the angular position on the sky of the five most probable CGs associated to each cluster with the whole BOSS spectroscopic sample, and identify those objects that are closer than 0.50 arcseconds, finding 3,772 matches for the most probable CGs that did not have an spectroscopic redshift measurement before. More than 99\% of all the identifications are done for pairs that are closer than 0.02 arcseconds, but we actually find a gap between 0.40 and 1.50 arcseconds where no identification at all is made, making clear that we are safe from any possible misidentification due to close galaxy pairs. We repeat this identification process with the second, third, fourth and fifth most probable CGs in each cluster. A sizeable fraction of CGs do not have measured redshifts yet whereas one or more of the highly probably CGs often does. This fuller comparison of members with redshifts provides now for the first time a highly spectroscopically complete sample of clusters with which make several precise calculations below described.

Finally, we exclude those clusters that, as can be seen in Fig.~\ref{fig:redMaPPer_Sky_Distribution}, are outside the considered BOSS area. This leaves us a final sample of 19,473 clusters with spectroscopic redshift measurement of their CG, and a total of 23,135 clusters with spectroscopic redshift measurement of one of their most likely centrals. In comparison, in the original photometric redshift catalogue there were 24,869 clusters in the same BOSS region, meaning that we have now spectroscopic information of more than the 93 percent of the clusters.

\begin{figure}
\resizebox{84mm}{!}{\includegraphics{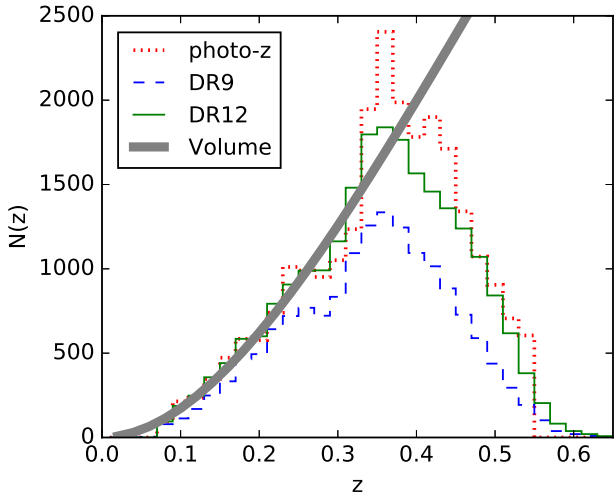}}
\caption{
Original photometric redshift (dot-dashed red line), DR9 spectroscopic redshift (dashed blue line) and updated DR12 spectroscopic redshift (solid green line) redMaPPer clusters redshift distribution. As the grey shaded line, we plot for comparison the proportional volume in each redshift bin. From the original 24,869 clusters with photometric redshift estimates, 15,936 clusters also have DR9 spectroscopic redshift information of their most likely CG, and 23,135 clusters have DR12 spectroscopic redshift measurements of at least one of their most likely CG. The clusters that fall outside the BOSS region have been excluded from these samples. Note that there is near full completeness in spectroscopic redshift to the peak of the redMaPPer selection function at $z=0.325$, and the bump feature at $z\simeq 0.25$ in the photometric redshift distribution, in contrast to the spectroscopic sample where the numbers smoothly rise with the increasing volume, as expected for a complete sample.}
\label{fig:redMaPPer_Redshift_Distribution}
\end{figure}

\begin{figure}
\resizebox{84mm}{!}{\includegraphics{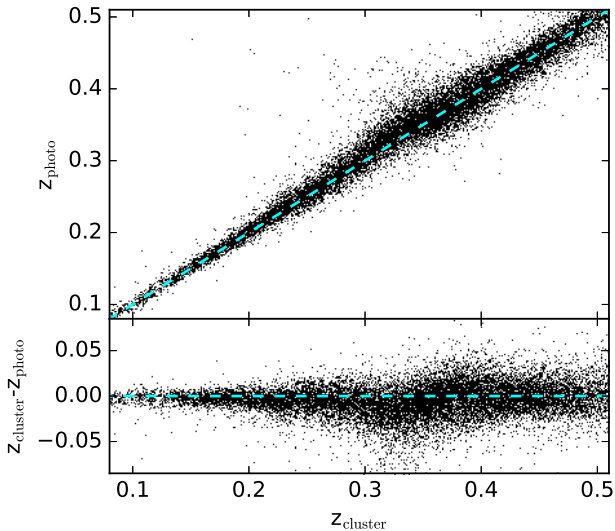}}
\caption{
Relation between our definition of $z_{\rm cluster}$ (Eq.~\ref{eq:z_cluster}), based on DR12 data and the centering probabilities $p_{{\rm cen},i}$ of all the CG candidates, and $z_{\rm photo}$ for each of the clusters in the redMaPPer catalogue. The residual wave behaviour is understood to be related to the ``quantization'' of photometric redshift estimates from the discrete wavelength coverage of the photometric bands.}
\label{fig:Spec_vs_Photo}
\end{figure}

To evaluate the final redshift of those clusters with more than one CG with a spectroscopic measurement, we make an average of their redshifts $z_{{\rm CG},i}$, weighted by the centering probability $p_{{\rm cen},i}$ associated to each of the five potential CGs:
\begin{equation}\label{eq:z_cluster}
z_{\rm cluster}=\frac{\sum\limits_{i=1}^5 z_{{\rm CG},i}\,p_{{\rm cen},i}}{\sum\limits_{i=1}^5 p_{{\rm cen},i}}
\end{equation}
The final redshift distribution of the clusters inside the BOSS region is shown in Fig.~\ref{fig:redMaPPer_Redshift_Distribution}. The difference between our definition of $z_{\rm cluster}$, and $z_{\rm photo}$, the original photometric redshift provided by the catalogue is shown in Fig.~\ref{fig:Spec_vs_Photo}.

%%%%%%%%%%%%%%%%%%%%%%%%%%%%%%%%%%%%%%%%%%%%%%%%%%%%%%%%%%%%%%%%%%%%%%%%%%%%%%%
%%%%%%%%%%%%%%%%%%%%%%%%%%%%%%%%%%%%%%%%%%%%%%%%%%%%%%%%%%%%%%%%%%%%%%%%%%%%%%%
\section{MXXL simulations}
\label{sect:mxxl}
%%%%%%%%%%%%%%%%%%%%%%%%%%%%%%%%%%%%%%%%%%%%%%%%%%%%%%%%%%%%%%%%%%%%%%%%%%%%%%%

The ``Millennium-XXL'' (MXXL, \citealp{Angulo2012}) simulation is one of the largest dark matter N-body simulations performed to date. It follows the nonlinear creation and growth of dark matter structures within a cube of 3,000 Mpc $h^{-1}$ on a side, which contained 6720$^3$ particles of mass $m_p = 8.456\times 10^9\,M_\odot$. 

Compared to its predecessor the ``Millennium Simulation'' (MS, \citealp{Springel2005a}), the MXXL simulation volume is 200 times bigger, comprising the equivalent volume of the Universe to $z=0.72$, or 7 times the volume of the BOSS survey. Although the particle resolution is 7 times lower in the MXXL simulation than in the MS, it is 300 times higher than the ``Hubble Volume Simulation'' \citep{Evrard2002} and has ample resolution for our massive clusters related purposes. Note that these large simulations deliberately share the same pre-{\em Planck} cosmology set by WMAP in 2003 to allow consistency checks. This is not in practice a limita-
tion for our work given the realization by \cite{Angulo2010} that a simple rescaling of size and redshifts can effectively provide predictions for other cosmologies using the same simulation, within the context of $\lm$.

This simulation was designed to provide enough statistical power as to study and interpret some of the problems related to the observation of galaxy clusters, like the scaling between the real mass of a cluster with the associated cluster observables, i.e., richness, lensing mass, X-ray luminosity and thermal Sunyaev-Zel'dovich signal. The code employed in the simulation is a extremely memory-efficient version of \texttt{GADGET-3}, which is itself a more sophisticated and efficient version of \texttt{GADGET-2} \citep{Springel2005b}, the code used for the MS.

This code is also designed to carry out halo and subhalo finding procedures during its execution with a friends-of-friends algorithm (FoF, \citealt{Knebe2011}), combined with the \texttt{SUBFIND} algorithm \citep{Springel2001}, which identifies locally overdense regions within the parent haloes found by the FoF. The halo catalogues produced during this search provide information about masses, velocity dispersions, halo shapes, velocities, etc...

The original cosmology employed in the MXXL simulation is $\lm$, with the same cosmological parameters that were employed in the MS: $\Omega_m=0.25$, $\Omega_\Lambda=0.75$, $\sigma_8=0.9$ and $H_0=73$ km s$^{-1}$ Mpc$^{-1}$. We update this cosmology and the resulting halo catalogue using the prescription given in \cite{Angulo2010}, with this new set of cosmological parameters taken from the combined analysis of {\em Planck} CMB data, BAO surveys and the JLA sample of Type Ia SNe \citep{Planck2015XIII}: $\Omega_m=0.3089$, $\Omega_\Lambda=0.6911$, $\sigma_8=0.8159$ and $H_0=67.74$ km s$^{-1}$ Mpc$^{-1}$. This algorithm to update simulations relies basically on the reassignments of masses, velocities and lengths, and the rescaling of the time steps, i.e., the redshifts of the snapshots, to match the shape of the smoothed linear matter power spectra of the desired cosmology, and thus the growth of structure. The modification of the long wavelength modes relies on the Zel'dovich approximation so the difference on large scales is also taken into account. We use the code \texttt{CAMB} \citep{Lewis2000} to compute the linear matter power spectra required for comparison. From now on, we will refer to this updated version as the MXXL simulations.

\subsection{MXXL synthetic cluster catalogue}

We now proceed to create a synthetic optical cluster catalogue from the MXXL simulations, so we can compare it with the redMaPPer catalogue. We use the data coming from 5 adjacent snapshots of the MXXL simulations, corresponding to the redshifts $z=0.027,\,0.128,\,0.242,\,0.393$ and $0.486$. 
%Old ones:$z=0.000,\,0.116,\,0.242,\,0.408$ and $0.509$. 
In each of these snapshots, we select of the order of $\sim7$ million DM haloes with masses above $M_{200c}=10^{12}\cmass$, where here $M_{200c}$ is defined as the mass that is enclosed in a sphere centered in the potential minimum of the halo, that has a mean density 200 times the critical density of the Universe.

Now, in order to build a synthetic cluster catalogue from the DM halo catalogue, where the ``true'' mass, the position and peculiar velocity of the halo is known, we need to assign a richness value and artificially place a CG in each of these haloes, which is, at the end, what is provided by the redMaPPer catalogue, as described in the previous section.

\subsubsection{Richness}

First of all, we need to associate an estimate of the richness to all the haloes in the MXXL simulations. We employ the usual mass--richness $\left<\MR\right>$ relation, which adopts the form:
\begin{equation}\label{eq:MR_relation}
 \ln \left( \frac{M_{200c}(\lambda)}{10^{14}\cmass} \right) = \ln \left( \frac{M_{200c}(\lambda_0)}{10^{14}\cmass} \right) + \alpha_{\MR}\,\ln\left(\frac{\lambda}{\lambda_0}\right)\,,
\end{equation}
where $M_{200c}(\lambda_0)$ is a reference mass at a given value of $\lambda=\lambda_0$, and $\alpha_{M|\lambda}$ is the slope of the mass--richness relation.

This relation has been the focus of intense research in the past decade, as an accurate scaling between the mass of a halo and the cluster observables is mandatory to constrain cosmological parameters from cluster counting techniques \citep{Rozo2007,Tinker2012}. If one wants to obtain a calibration which does not rely on matching a certain cosmology, a relation between the richness $\lambda$ and the lensing mass $M_{\rm lens}$, the X-ray luminosity $L_X$, or the thermal Sunyaev-Zel'dovich signal $Y_{SZ}$ must be found. \cite{Johnston2007}, using weak lensing measurements, find a slope of $\alpha_{\MR}=1.28 \pm 0.04$ for a reference mass of $M_{200}(\lambda_0=20) = 0.88 \times10^{14} \cmass$. \cite{Wen2010}, using a sample of 24 nearby clusters with weak lensing and X-ray mass measurements, found a steeper slope $\alpha_{\MR}=1.55$, although it should be noted that their richness definition differs from the one given by maxBCG and redMaPPer algorithms. More recently, using SDSS weak lensing data and the redMaPPer catalogue in the $0.1\leqslant z \leqslant 0.33$ redshift region, \cite{Simet2016} find a slope of $\alpha_{\MR}=1.33^{+0.09}_{-0.10}$ for a reference mass of $M_{200m}(\lambda_0=40) \sim 2.22 \times10^{14} \cmass$. On the other hand, one can assume a determined cosmology and use matching abundances techniques to constrain this scaling relation. \cite{Rykoff2012} provide a tentative mass--richness relation with an slope of $\alpha_{\MR}=1.06$ for a reference mass of $M_{200c}(\lambda_0=60) = 4.4\times10^{14}\cmass$. For a fixed cosmology and using the photometric redshift estimates from DR8, \cite{Baxter2016} measured the angular clustering of redMaPPer clusters in order to calibrate the mass--richness and mass--bias relations, and obtained an slope of $\alpha_{\MR}=1.18\pm0.16$ and no indication of any evolution of the mass--richness relation with redshift.

Because these cluster observables are intrinsically noisy, for each mass-observable relation there is a scatter associated to the distribution of the values of such observable given a true mass. The uncertainty in the value of this scatter alters the effectivity of any cosmological analysis as much as the uncertainty in the mean mass--observable relation. The fractional scatter $\sigma_{\MR}$ on the halo mass at fixed richness can be described by a log-normal distribution:
\begin{equation}
\Delta \ln \left( \frac{M_{200c}(\lambda)}{10^{14}\cmass} \right) = \sigma_{\MR}\,.
\end{equation}
There have been several studies in the past years trying to evaluate $\sigma_{\MR}$, using either simulations or data, considering mass proxies like X-ray luminosity.
\cite{Rozo2009a}, using maxBCG catalogue and weak lensing and X-ray data, found $\sigma_{\MR} = 0.45^{+0.20}_{-0.18}$ at $\lambda \approx 40$. 
Introducing modifications on the maxBCG richness estimator (modifications that would later be implemented in the redMaPPer algorithm), \cite{Rykoff2012} obtained a value of the scatter at fixed mass between $\sigma_{\MR} \approx 0.2-0.3$.
Taking a different approach, \cite{Angulo2012} first populated the MXXL simulations haloes with galaxies using a halo occupation distribution (HOD) model, identified clusters and measured their richness using a procedure similar to the one employed by the maxBCG algorithm, and obtained a predicted scatter of $\sigma_{\MR} = 0.36$.
\cite{Becker2007} found relatively higher values for the scatter, $\sigma_{\MR} \sim 0.6$, using the velocity dispersion in maxBCG clusters to obtain a mass estimation. 

\begin{figure}
\resizebox{84mm}{!}{\includegraphics{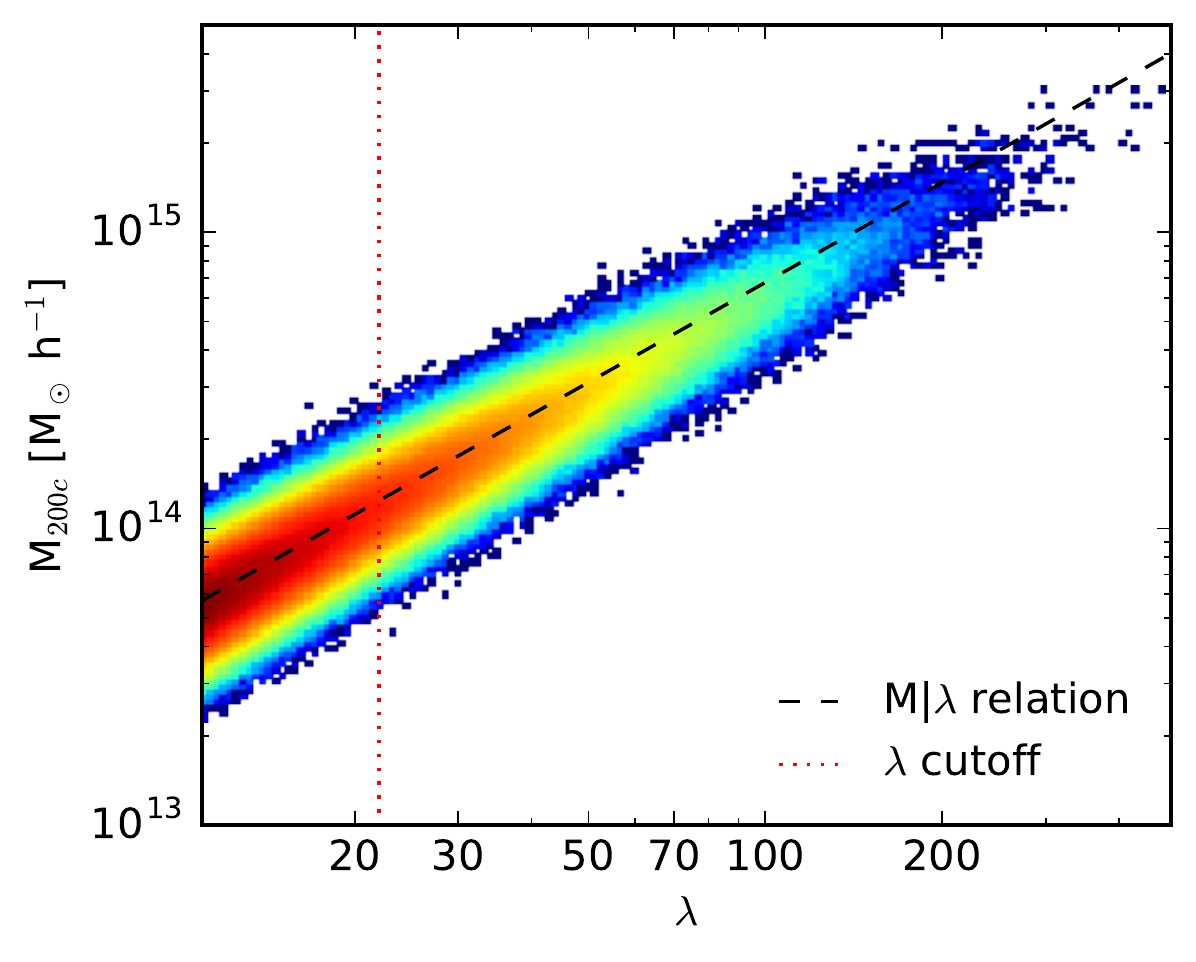}}
\caption{Mass distribution as a function of richness of the MXXL DM haloes found in snapshot 54, that corresponds to $z=0.242$, close to the mean redshift of the complete cluster sample. The ``true'' distribution, shown as the black dashed line, is the theoretical mass--richness relation with $\kappa_{\MR}=1.35$ and $\alpha_{M|\lambda} = 1.10$. When the scatter is not considered, there are $\sim$116,000 clusters above the richness threshold. If a scatter of $\sigma_{M|\lambda}=0.20$ is introduced, we obtain the cluster distribution shown in the figure, with around $\sim$128,000 clusters with richness, $\lambda>22$ due to the upscatter of low mass clusters.}
\label{fig:MR_relation}
\end{figure}

In Sec.~\ref{sect:likelihood}, we will find, through a likelihood analysis, the values of $\kappa_{\MR}$, $\alpha_{\MR}$ and $\sigma_{\MR}$ that best describe the observations, where for clarity we have defined:
\begin{equation}
\kappa_{\MR} \equiv \ln \left( \frac{M_{200c}(\lambda_0=60)}{10^{14}\cmass} \right)\,,
\end{equation}
and use these values in Secs.~\ref{sect:correlation} and \ref{sect:abundances} when comparing any data obtained from the redMaPPer with the ``model'' produced by the MXXL simulations. We will follow the results obtained by \cite{Angulo2012} and \cite{Rozo2010}, and consider the value of the scatter independent of the richness, and ignore any possible redshift evolution of the scatter or of the slope of the mass--richness relation. Also, because in the simulations we know the value of the true mass, rather than the value of the observable, we need to convert $\sigma_{\MR}$ into $\sigma_{\lambda|M}$ inverting Eq.~\ref{eq:MR_relation}, so $\sigma_{\MR}= \alpha_{\MR}\,\sigma_{\lambda|M}$. In Fig.~\ref{fig:MR_relation} we show the distribution of the masses of the MXXL clusters as a function of one realization of the richness associated through a mass--richness relation with, e.g., a pivot mass of $\kappa_{\MR}=1.35$, a slope of $\alpha_{M|\lambda} = 1.10$, and a scatter of $\sigma_{\MR} = 0.20$, in comparison to the ideal mass--richness relation without scatter. The upscattering of low mass clusters into high richness regions increases with higher values of the slope or larger scatter.

\subsubsection{CG miscentering}

Recent studies have shown that the assumption that central galaxies (CGs) lie basically at rest in the deepest part of the potential well is not accurate. Although there is evidence of an special correlation between the mass of a cluster halo and the properties (mass, morphology, star formation rates, stellar population, colour, etc...) of its brightest cluster galaxy (BCG, \citealp{vonderLinden2007}), we may expect a large proportion of clusters are non-relaxed dynamical systems that are still evolving, with BCGs following an evolving orbit that need not be located at the time varying minimum of the cluster potential.

Miscentering is one of the main sources of error on stacked measurements on clusters, including weak-lensing mass determinations \citep{Umetsu2011b, Umetsu2011a}, analyses of the power spectrum \citep{Reid2010}, gravitational redshift measurements \citep{Jimeno2015}, or velocity dispersion calculations \citep{Becker2007}. That is why in the last few years many authors have tried to determine the level of miscentering statistically.

There are two reasons why this miscentering may occur: the galaxy identified as the CG not being the galaxy with the lowest specific potential energy (i.e., the one that could be considered as the CG), or the real CG not being in the center of mass of the DM halo.

In cluster catalogues like the maxBCG \citep{Koester2007b}, the GMBCG \citep{Hao2010} or the WHL12 \citep{Wen2012}, the center of a cluster is identified with the position of the BCG. Although this assumption seems to improve the overall centering performance of cluster-finder algorithms \citep{Rozo2014a}, \cite{Skibba2011} claimed that between the 25\% and the 40\% of cases the BCG is not the real CG, but a satellite galaxy, and \cite{vonderLinden2007} found in a sample of 625 clusters that in more than a half of them the BCG was not located in the center of the cluster.

Trying to quantify the level of miscentering, and using mock catalogues, \cite{Soares-Santos2011} found an offset distribution that could be fitted by a 2D gaussian with a standard deviation of $\sigma=0.47\hdist$, meanwhile \cite{Johnston2007} found $\sigma=0.42\hdist$ for the BCGs that were not accurately centered, which ranged from 40\% to 20\% as a function of cluster richness.

In order to address this problem, the redMaPPer iterative self-training centering algorithm used BCGs as the seeds for the centering process, but in successive calibrating iterations all the galaxies found in the cluster that were consistent with the red-sequence were considered as potential CGs, and had a centering probability assigned. When convergence is obtained, the galaxy with the highest probability of being in the center of the cluster is tagged as the CG. It should be noted that the miscentering introduced by the red-sequence prior that does not allow blue galaxies to be selected as CGs is expected to affect less than 2\% of the clusters. In any case, when the cluster centers found on high resolution X-ray data were compared to the CGs found in the redMaPPer catalogue, the redMaPPer algorithm was claimed to have a centering success rate of $\approx 86\%$ \citep{Rozo2014a}. 

To include this effect into our synthetic catalogue, we follow the results of \cite{Johnston2007} and introduce a miscentering probability as a function of the halo richness. The probability $p_{mc}$ of a CG being displaced from the center of its host halo follows: $p_{mc}(\lambda) = (2.13+0.046\,\lambda)^{-1}$, and such displacement is given by a 2D Gaussian of width $\sigma_{mc}=0.42\hdist$.

\subsubsection{CG peculiar motion}

There are good reasons to believe that, compared to other member galaxies, CGs are a cold population of galaxies due to dynamical friction or possible central gas cooling. How much so is still a matter of discussion. Hierarchical merging of clusters means we must expect some level of dispersion periodically as CGs respond to a rapidly evolving potential and merge with each other. In any case, it is clear that their peculiar velocity cannot be ignored and in some cases is very anomalous \citep{Sharples1988} and with a small net gravitational redshift expected \citep{Broadhurst2000} and measured \citep{Wojtak2011, Jimeno2015, Sadeh2015}. Puzzlingly, a significant fraction of CGs with peculiar motions are located in the peak of the X-ray emission. Given the hydrodynamical forces relevant for the cluster, it is not expected that the gas should move together with the CG galaxy during cluster encounters \citep{Ricker2001, Molnar2015}.

The relation between the value of $\sigma_{\rm CG}$, the distribution of CG motions, and $\sigma_{\rm gal}$, the value of the dispersion associated to the cluster satellite galaxies, is still quite unknown. \cite{Oegerle2001} found that from a sample of 25 clusters almost all CGs showed peculiar velocities relative to the mean velocity of the clusters studied, with $\sigma_{\rm BCG}\approx175\kms$. \cite{Coziol2009}, studying a much larger sample of clusters, found that CGs having peculiar motions within the cluster was a general phenomenon, with less than the 30 percent of them having velocities compatible with zero, and more than half of them having velocities higher than $0.3\,\sigma_{\rm gal}$, depending this value slightly on cluster richness. \cite{Skibba2011}, studying the miscentering of CGs, also found a relatively high value for their velocities, with $\sigma_{\rm CG}\simeq0.5\,\sigma_{\rm gal}$, that had also little dependence with the mass of the host cluster. 

In any case, to mimic this motion within the cluster of our already placed CGs, we will assign a peculiar velocity to them given by $\sigma_{\rm CG} \simeq 0.4 \, \sigma_{\rm vir}$, where $\sigma_{\rm vir}$ is the virial velocity associated to the mass of the cluster.

Once this CG peculiar motion has been added to the motion of the cluster, we move CGs from real-space positions $\bm{x}$ to redshift-space positions $\bm{s}$:
\begin{equation}
\bm{s} = \bm{x} + \left( 1 + z \right) \frac{\bm{v} \cdot \bm{\hat{x}_i}}{H(z)}\,\bm{\hat{x}_i}\,,
\end{equation}
where $\bm{\hat{x}_i}$ is an arbitrarily line-of-sight chosen direction, and $\bm{v}$ is the final velocity of the CG.

%%%%%%%%%%%%%%%%%%%%%%%%%%%%%%%%%%%%%%%%%%%%%%%%%%%%%%%%%%%%%%%%%%%%%%%%%%%%%%%
%%%%%%%%%%%%%%%%%%%%%%%%%%%%%%%%%%%%%%%%%%%%%%%%%%%%%%%%%%%%%%%%%%%%%%%%%%%%%%%
\section{Correlation Function}
\label{sect:correlation}
%%%%%%%%%%%%%%%%%%%%%%%%%%%%%%%%%%%%%%%%%%%%%%%%%%%%%%%%%%%%%%%%%%%%%%%%%%%%%%%

We compute the two-point redshift-space correlation function $\xi(s,\mu)$ using the \cite{Landy1993} estimator, where $s$ is the redshift-space distance in $\hdist$ units, and $\mu = \cos\theta$, where $\theta$ is the angle of the pair with respect to the line-of-sight:
\begin{equation}
 \xi(s,\mu)=\frac{DD(s,\mu)-2DR(s,\mu)+RR(s,\mu)}{RR(s,\mu)}\,,
\end{equation}
where $DD(s,\mu)$, $DR(s,\mu)$ and $RR(s,\mu)$ are the normalized number of pairs found at a distance of $s$ and an angular separation of $\mu$ in the data-data, data-random and random-random samples.

We want to investigate the multipoles of the correlation function $\xi(s,\mu)$, which are:
\begin{equation}
 \xi_\ell(s)=\left(\frac{2\ell+1}{2}\right)\int_{-1}^{+1}\xi(s,\mu)L_\ell(\mu)d\mu\,,
\end{equation}
where $L_\ell(\mu)$ is the $\ell$th Legendre polynomial. We weight $DD$, $DR$ and $RR$ pairs with the associated value of $L_\ell(\mu)$ for the monopole ($\ell=0$) and the quadrupole ($\ell=2$):
\begin{equation}
 \xi_\ell(s)=\left(\frac{2\ell+1}{2}\right)\frac{DD_\ell(s)-2DR_\ell(s)+RR_\ell(s)}{RR_0(s)}
\end{equation}

To optimally weight regions with different number densities, we apply FKP weighting \citep{Feldman1994} to each cluster:
\begin{equation}
 w_P=\frac{1}{1+n(z)P_{_{\rm FKP}}}\,,
\end{equation}
where $n(z)$ is the mean cluster density at redshift $z$, and $P_{_{\rm FKP}}=$ 20,000$ \,h^3\,{\rm Mpc}^{-3}$.

For the data sample, i.e., redMaPPer clusters, instead of considering only the most probable CGs, we use all the five CG candidates provided in the redMaPPer catalogue for each cluster, taking their centering probabilities $p$ to weight their contribution to the final pair count, so that $\sum_{\alpha=1}^5 p_{i,\alpha} = 1$ for cluster $i$. The final data-data, data-random and random-random pair counts can be expressed as:
\begin{equation}
DD_\ell(s)=\sum_{i=1}^N \sum_{j=i+1}^N \sum_{\alpha=1}^5 \sum_{\beta=1}^5 w_{ P,i}\,w_{P,j}\,p_{i,\alpha}\,p_{j,\beta}\,L_\ell(\mu)\,,
\end{equation}
\begin{equation}
DR_\ell(s)=\sum_{i=1}^N \sum_{j=1}^{\widetilde{N}} \sum_{\alpha=1}^5 w_{ P,i}\,w_{P,j}\,p_{i,\alpha}\,L_\ell(\mu)\,,
\end{equation}
\begin{equation}
RR_\ell(s)=\sum_{i=1}^{\widetilde{N}} \sum_{j=i+1}^{\widetilde{N}} w_{ P,i}\,w_{P,j}\,L_\ell(\mu)\,,
\end{equation}
where $N$ is the number of clusters in the data sample, and $\widetilde{N}$ is the number of objects in the synthetic random catalogue.

We are also interested in the projected correlation function $\Xi(r_\perp)$, which provides information of the real-space clustering so that we do not need to worry about the complications of peculiar motions \citep{Davis1983}. It is obtained integrating the 2D correlation function $\xi(r_{los},\,r_{\perp})$ along the line-of-sight:
\begin{equation}
\Xi(r_\perp)=2\int_0^\infty \xi(r_{los},\,r_{\perp})\,dr_{los}\,,
\end{equation}
where $r_{los}=\mu\,s$, and $r_\perp=\sqrt{s^2-r_{los}^2}$, are the parallel and perpendicular directions to the line-of-sight. To compute it from $\Xi(r_\perp)$, we use the estimator:
\begin{equation}
\Xi(r_\perp)=2\sum_i^{r_{los\,\max}} \xi(r_{los},\,r_{\perp i})\,\Delta r_{los}\,,
\end{equation}
where we bin the 2D correlation function into linearly spaced bins of constant size $\Delta r_{los} = 5\hdist$, and select a maximum summation distance of $r_{los\,\max}=30\hdist$.

We use the jackknife method to compute the covariance matrices of the correlation function. For each cluster sample we randomly create 80 cluster subsamples that comprise 1/80th part of the total, and then compute 80 times the monopole, dipole and projected correlation function of the total cluster sample with one of those cluster subsamples removed. The covariance matrix associated to this sample is then:
\begin{equation}
C_{i,j}=\frac{N-1}{N}\sum_{k=1}^N \left(\overline{\chi}_i - \chi_i^k\right) \left(\overline{\chi}_j - \chi_j^k\right)\,,
\end{equation}
where $\chi_i$ corresponds to $\xi_0$, $\xi_2$ or $\Xi$ at the $i^{th}$ bin, and $\overline{\chi}_i$ is the mean value of the $N=80$ calculations at the $i^{th}$ bin.

\subsection{Results}

\subsubsection{Redshift-space two-point correlation function}

\begin{figure}
\resizebox{84mm}{!}{\includegraphics{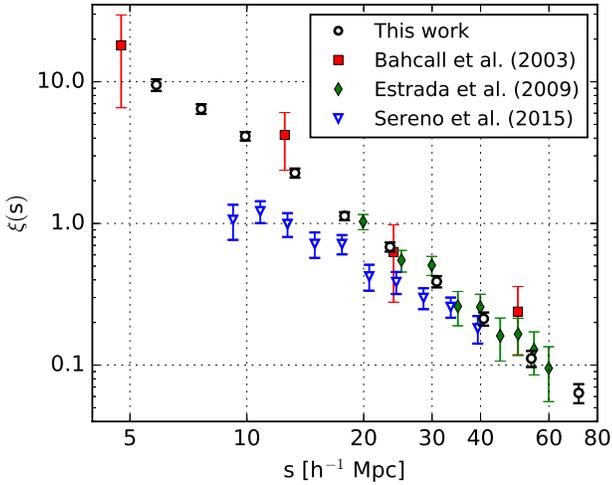}}
\caption{Our measurement of the redshift space correlation function (black circles) shows the huge improvement in precision now possible with the latest SDSS release, for the spectroscopically complete sample of 7,143 redMaPPer clusters contained in the full range $0.08 \leqslant z \leqslant 0.325$, with a lower richness limit of $\lambda>22$. For comparison, we plot previous measurements found in the literature, using other SDSS cluster samples: Bahcall et al. (2003) as red squares, Estrada et al. (2009) as green diamonds, and Sereno et al. (2015b) as blue triangles. Notice the correlation function slope and amplitude is similar to previous work but with much higher precision because of the larger number of clusters sampled to higher redshift, and without the photometric redshifts that smooth the Sereno et al. (2015b) correlations (blue triangles) at small scales.}
\label{fig:Comparison_CF}
\end{figure}

First of all, and as a check of the power of both the spectroscopic sample based on the redMaPPer catalogue that we have constructed, and our probability-weighted estimator $\xi(s)$, we compute the redshift-space two-point correlation function up to $s=80\hdist$. For that, we use a test sample containing all the clusters in the $0.080 \leqslant z \leqslant 0.325$ redshift range, and richness $\lambda>22$, which contains 7,143 clusters. The values obtained are shown in Fig.~\ref{fig:Comparison_CF}, where we also compare our results with previous measurements found in the literature \citep{Bahcall2003, Estrada2009, Sereno2015b}. Notice in Fig.~\ref{fig:Comparison_CF} the correlation function rises continuously to small radius with a slope similar to previous work but with much higher precision because of the larger numbers of clusters sampled to higher redshift. The increasingly shallower slope of \cite{Sereno2015b} at smaller scales is due to smoothing by the relatively large proportion of photometric redshifts in their analysis of the GMBCG cluster catalogue.

\subsubsection{Measured monopole, quadrupole, 2D, and projected correlation function}

Here we divide the redMaPPer catalogue into three different richness bins, $\lambda_1\in[22, 30)$, $\lambda_2\in[30, 45)$, and $\lambda_3\in[45, 200)$, with similar numbers of clusters and each of the resulting subsamples is again divided into two redshift regions in order to have a ``low-z'' and a ``high-z'' sample so we can examine evolution. The redshift regions for each of these six subsamples, together with the mean redshift, mean richness and number of clusters contained, are listed in Table~\ref{tab:redMaPPer_Subsamples}.

%================================================
\begin{table}
\center
\caption{
\label{tab:redMaPPer_Subsamples}
Richness range $\lambda$, redshift region $z_{\rm clu}$, number of clusters $N$, mean richness $\left<\lambda\right>$, and mean redshift $\left<z_{\rm clu}\right>$ of the six redMaPPer cluster subsamples considered in the measurement of the correlation function.
}
\begin{tabular}{l c c c c c}
\hline
\hline
Subsample 		&$\lambda$ 	& $z_{\rm clu}$ 	& $N$	 	& $\left<\lambda\right>$ 	& $\left<z_{\rm clu}\right>$\\
\hline
$\lambda_1$ low-z	&$[22, 30)$	& $[0.080, 0.250)$	&1770 		&25.3 		&0.189  	\\
$\lambda_1$ high-z	&$[22, 30)$	& $[0.250, 0.400]$	&4493 		&25.6 		&0.334  	\\
\vspace{-5pt}\\
$\lambda_2$ low-z	&$[30, 45)$	& $[0.080, 0.275)$	&1527 		&36.0 		&0.205  	\\
$\lambda_2$ high-z	&$[30, 45)$	& $[0.275, 0.425]$	&4008 		&35.9 		&0.363  	\\
\vspace{-5pt}\\
$\lambda_3$ low-z	&$[45, 200)$	& $[0.080, 0.300)$	&1024 		&63.1 		&0.221  	\\
$\lambda_3$ high-z	&$[45, 200)$	& $[0.300, 0.450]$	&2384 		&62.11 		&0.388  	\\

\hline
\end{tabular}
\end{table}
%================================================

\begin{figure*}
 \centering
 \includegraphics[width=0.98\textwidth]{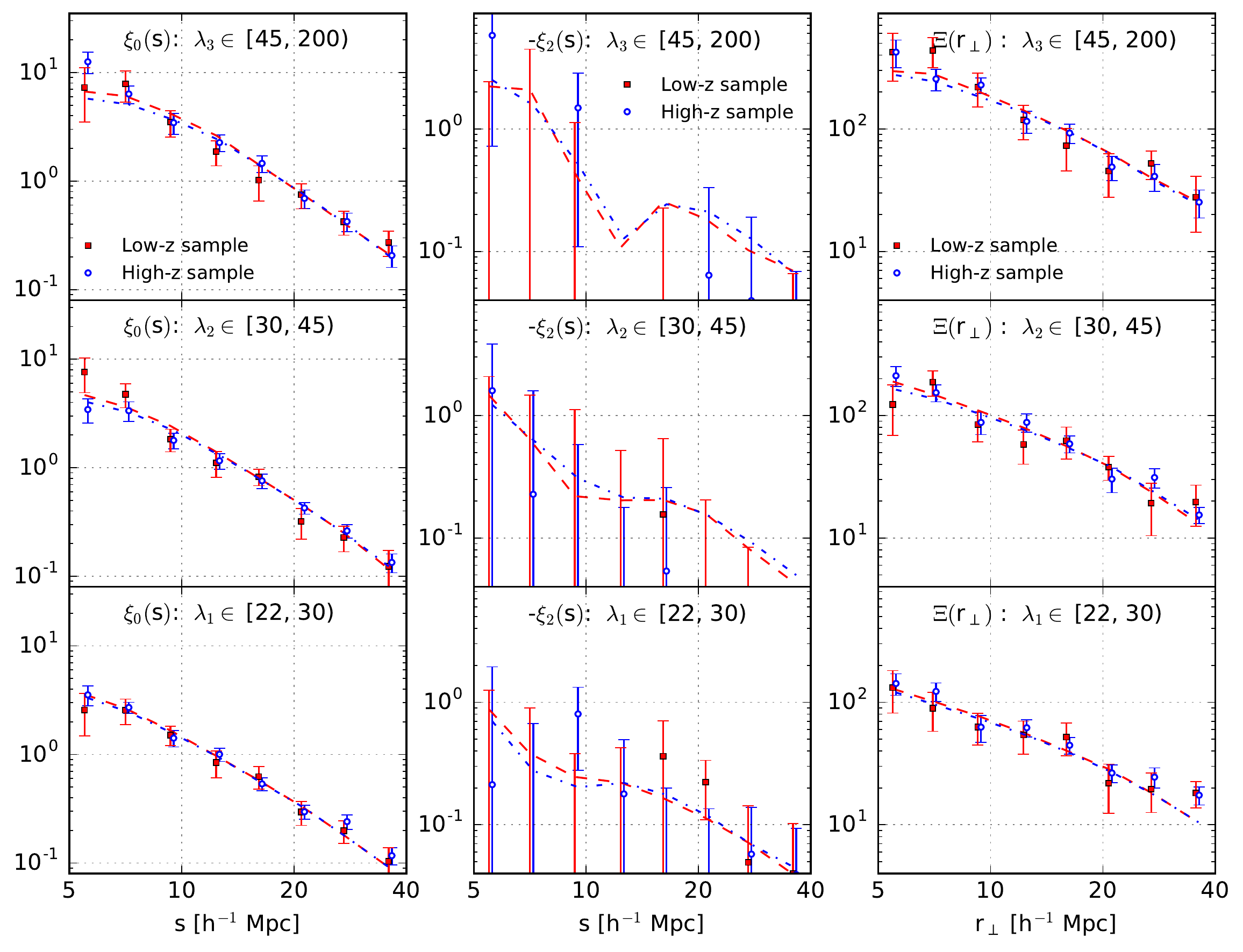}
 \caption{Monopole $\xi_0(s)$ (left column), quadrupole $\xi_2(s)$ (central column) and projected correlation function $\Xi(r_\perp)$ (right column) for two redshift samples: low-z (red squares for redMaPPer and red dashed lines for the model) and high-z (blue circles for redMaPPer and blue dot-dashed lines for the model), and three richness ranges: $\lambda \in [22, 30)$ (upper panels), $[30, 45)$ (central panels), and $[45, 200)$ (lower panels). Notice the clear increase of the amplitude of the correlation functions with richness, and the near independence with redshift, in excellent agreement with the model predictions, which match very well with radius and with richness. The model curve here is derived in Sec.~\ref{sect:likelihood} from our likelihood analysis based on the MXXL simulations, and the best fitting mass--richness relation.}
 \label{fig:1D_CF}
\end{figure*}

\begin{figure*}
 \centering
 \includegraphics[width=0.49\textwidth]{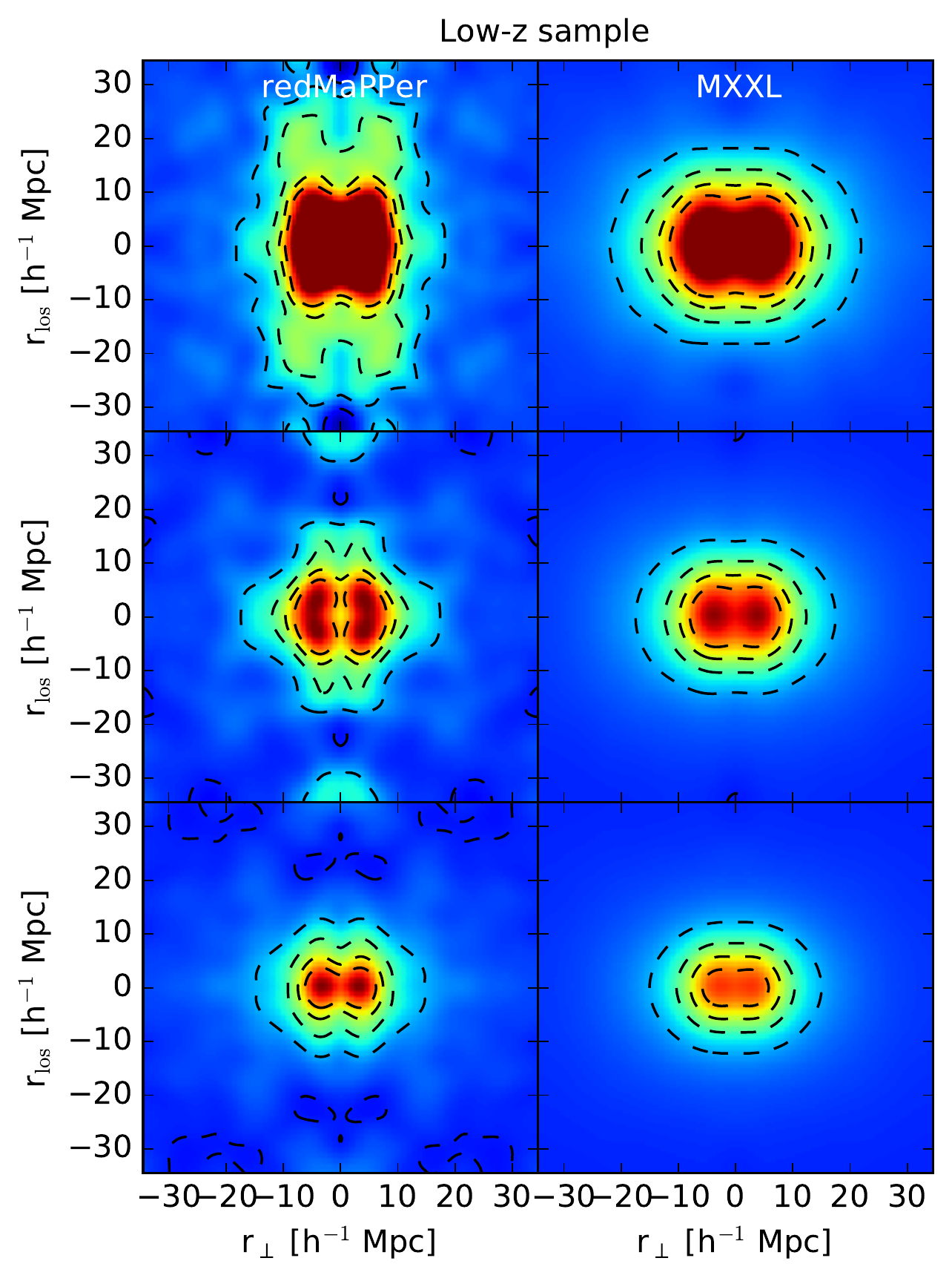}
 \includegraphics[width=0.453\textwidth]{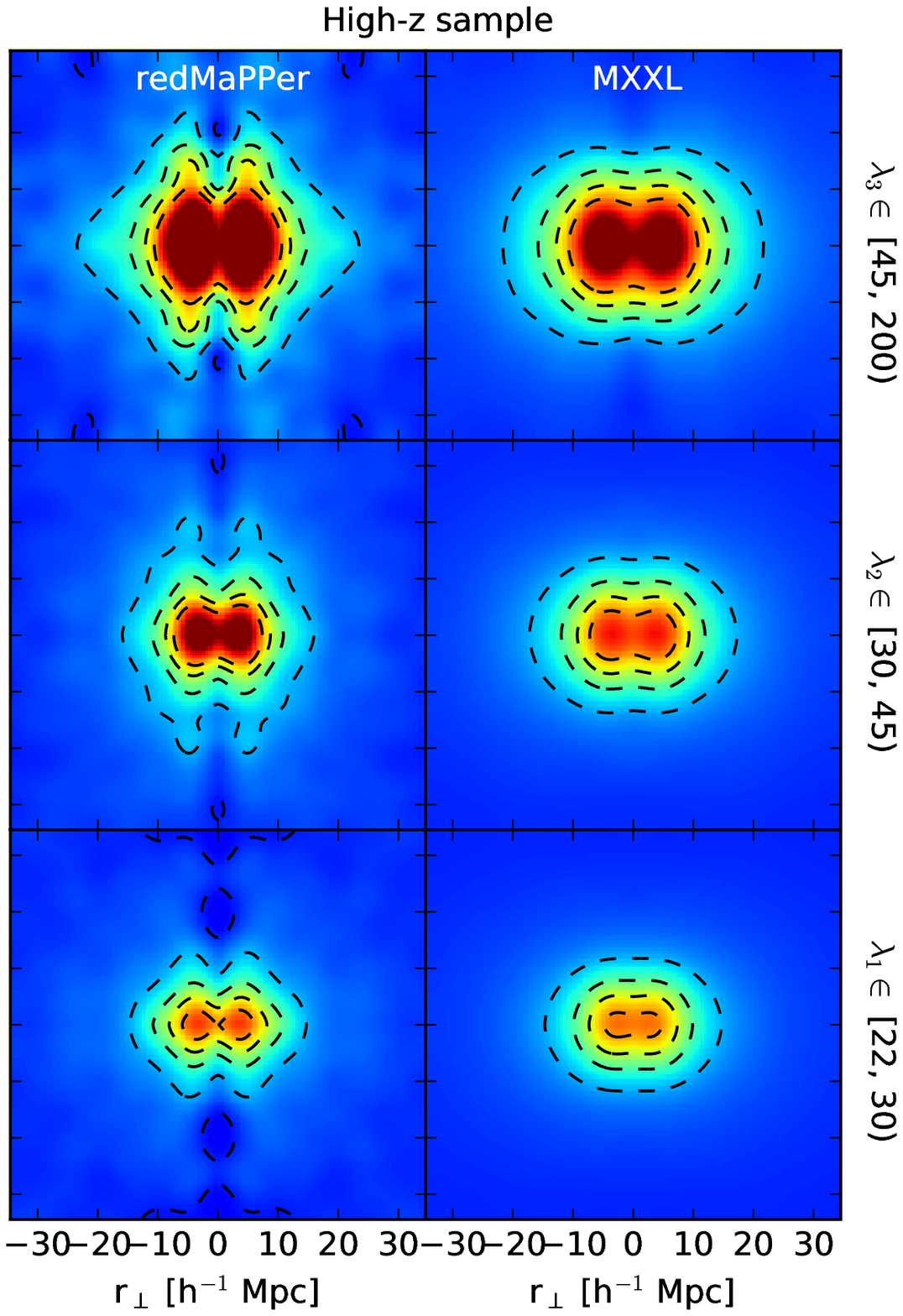}
\caption{2D correlation function $\xi(r_{los},\,r_{\perp})$ for two redshift samples: low-z (left panels) and high-z (right panels), and three richness bins: $\lambda \in [22, 30)$ (bottom), $[30, 45)$ (middle), and $[45, 200)$ (top). The model results derived in Sec.~\ref{sect:likelihood} from the MXXL simulations are shown for comparison. Dashed contours correspond to values $\xi(r_{los},\,r_{\perp}) = (0.0, 1.5, 3.0, 4.5)$. For clarity, a 2D gaussian smoothing with a kernel of width $5 \hdist$ has been applied to the images. Some significant differences are apparent here in the redshift direction given peculiar motions of the member galaxies that define the observed cluster redshifts.}
 \label{fig:2D_CF}
\end{figure*}

We now compute the monopole $\xi_0(s)$, the quadrupole $\xi_2(s)$, the 2D correlation function $\xi(r_{los},\,r_{\perp})$, and the projected correlation function $\Xi(r_\perp)$ following the procedure described before. We bin s in 8 logarithmic distributed bins between 5$\hdist$ and 35$\hdist$, and bin $r_{los}$ and $r_{\perp}$ in linearly spaced bins of a size equal to 0.5$\hdist$. The obtained values of $\xi_0$, $\xi_2$ and $\Xi$ for the six subsamples are shown in Fig~\ref{fig:1D_CF}, together with the MXXL realization model that best describes the real-space correlation function (see Sec.~\ref{sect:likelihood}). In the same way, the redMaPPer and the MXXL 2D correlation functions are shown in Fig.~\ref{fig:2D_CF}. We see a clear trend towards higher correlation amplitude with richness, and little dependence on redshift.
Later in Sec.~\ref{sect:likelihood} we discuss these measurements in comparison with the simulations predictions for the correlation functions after first defining the mass--richness relation between the mass function of the simulations with the observed cluster abundances.

Although the measurements of the quadrupole are too noisy to obtain any information from them, as we can see there is a clear increase in the amplitude of both the monopole and the projected correlation function for higher richness bins, following the behaviour found in the MXXL simulations. In each of these three subsamples, there is no clear evidence of any evolution between the two redshift bins, which, as we will see below is in good agreement with the MXXL simulations for the relatively small redshift range of the data. Some differences are apparent here in the redshift direction in Fig.~\ref{fig:2D_CF} with enhancement along the line of of sight compared to MXXL, which could be due to the higher than expected peculiar motion of the CGs used to define the observed cluster redshifts.

\subsubsection{Correlation length}

In order to fit the real-space correlation function $\xi(r)$, we approximate it by a power law:
\begin{equation}
\xi(r)= \left(\frac{r}{r_0}\right)^{-\gamma}\,,
\end{equation}
where both the correlation length $r_0$ and the slope $\gamma$ are left as free parameters. Considering this power law, the projected correlation function is:
\begin{equation}
\begin{split}
\Xi(r_\perp) & =2\int_0^\infty \xi\left[\left(r_\perp^2+r_{los}^2\right)^{1/2}\right]\,dr_{los} \\
& = 2\int_{r_\perp}^\infty \xi(r)\left(r^2-r_\perp^2\right)^{-1/2}\,dr \\
& = \sqrt{\pi}\,\frac{\Gamma\left(\left(\gamma-1\right)/\,2\right)}{\Gamma\left(\gamma\,/\,2\right)}\,r_0^\gamma\,r_\perp^{1-\gamma}\,.
\end{split}
\end{equation}

\begin{figure}
\resizebox{84mm}{!}{\includegraphics{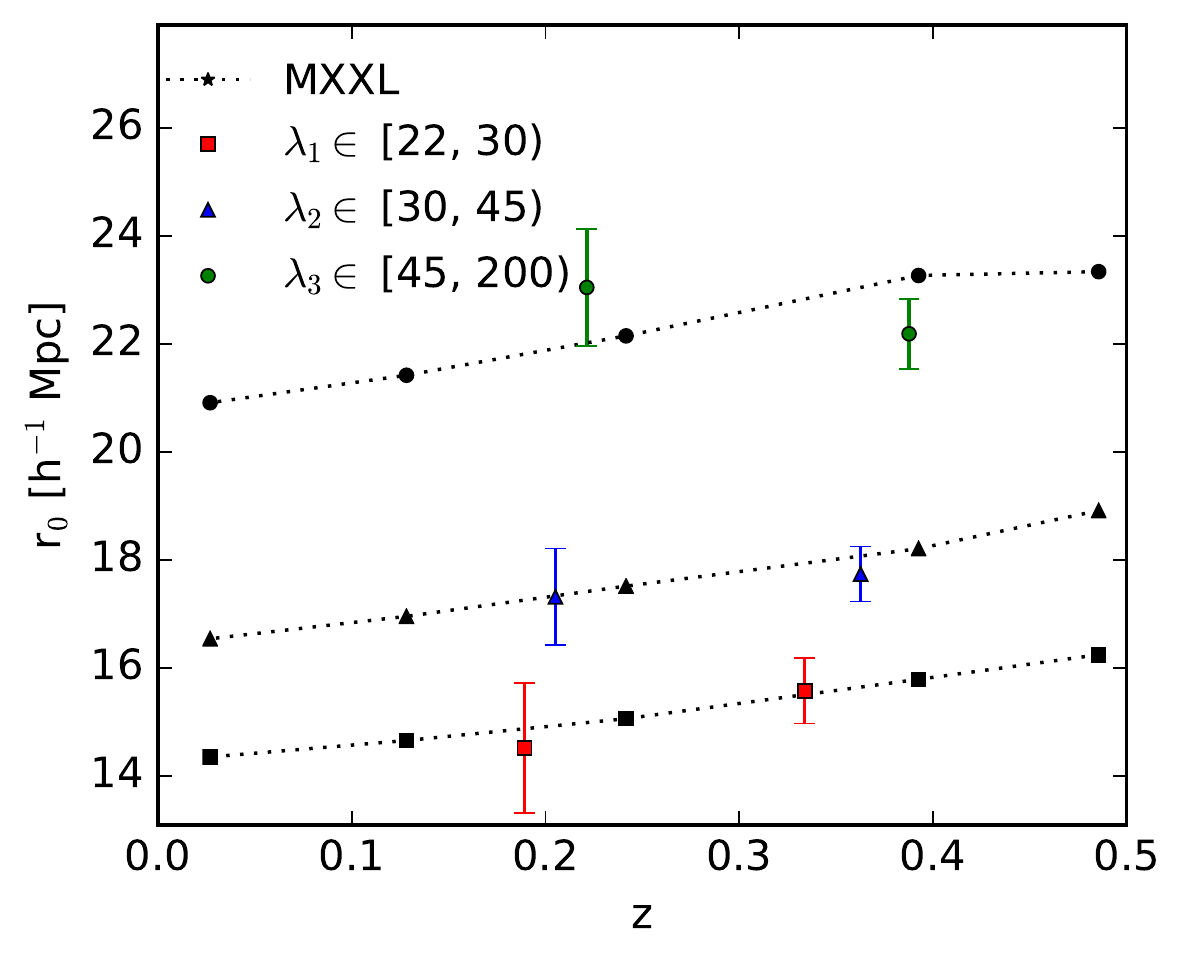}}
\caption{Values of the correlation length $r_0$ obtained for the six redMaPPer subsamples listed in Table~\ref{tab:redMaPPer_Subsamples} as a function of the average redshift of the sample. Within two redshift regions, three richness ranges have been considered: $\lambda \in [22, 30)$ (red squares), $[30, 45)$ (blue triangles), and $[45, 200)$ (green circles). Shown as dotted lines, the model values of $r_0$ that we obtain from the MXXL simulations for these three richness ranges (from bottom to top, respectively) and the snapshots available. The slow increasing trend with redshift that is apparent here for all three richness ranges is well matched by MXXL and corresponds to the increasing bias at fixed mass with redshift, corresponding to rarer more biased peaks in the density field (Kaiser 1986). The mass--richness relation needed to obtain the MXXL model curves is obtained through a likelihood analysis, as described later in Sec.~\ref{sect:likelihood}.}
\label{fig:Projected_CF_Fit_vs_z}
\end{figure}

The values found for both $r_0$ and $\gamma$ for each of the six redMaPPer subsamples considered before are given in Table~\ref{tab:redMaPPer_CF}. The redshift and richness dependency of these results is shown in Fig.~\ref{fig:Projected_CF_Fit_vs_z}, in comparison with the MXXL model that adopts the ``clustering'' mass--richness relation parameters described later in Sec.~\ref{sect:likelihood}.

%================================================
\begin{table}
\center
\caption{
\label{tab:redMaPPer_CF}
Values of the correlation length $r_0$ and the real-space correlation function slope $\gamma$ obtained for the six redMaPPer cluster subsamples considered in Table~\ref{tab:redMaPPer_Subsamples}.
}
\begin{tabular}{l c c}
\hline
\hline
Sample 			&$r_0\,[\hdist]$ 	&$\gamma$	\\
\hline
$\lambda_1$ low-z	&$14.53 \pm 1.20$	&$2.04 \pm 0.18$\\
$\lambda_1$ high-z	&$15.58 \pm 0.61$	&$2.16 \pm 0.11$\\
\vspace{-5pt}\\
$\lambda_2$ low-z	&$17.32 \pm 0.90$	&$2.26 \pm 0.19$\\
$\lambda_2$ high-z	&$17.74 \pm 0.51$	&$2.34 \pm 0.10$\\
\vspace{-5pt}\\
$\lambda_3$ low-z	&$23.05 \pm 1.08$	&$2.55 \pm 0.20$\\
$\lambda_3$ high-z	&$22.19 \pm 0.65$	&$2.52 \pm 0.13$\\
\hline
\end{tabular}
\end{table}
%================================================

The values of $r_0$ for the cluster subsamples considered agree, within the noise, with the MXXL simulations expected values, with the only exception of the $\lambda_3$ high-z subsample, which is slightly more than 1$\sigma$ below the expect value. This could be an indication of the limitations of the redMaPPer algorithm above $z>0.35$, where it may be overestimating the richness of some clusters and thus diluting the amplitude of the correlation function.

\begin{figure}
\resizebox{84mm}{!}{\includegraphics{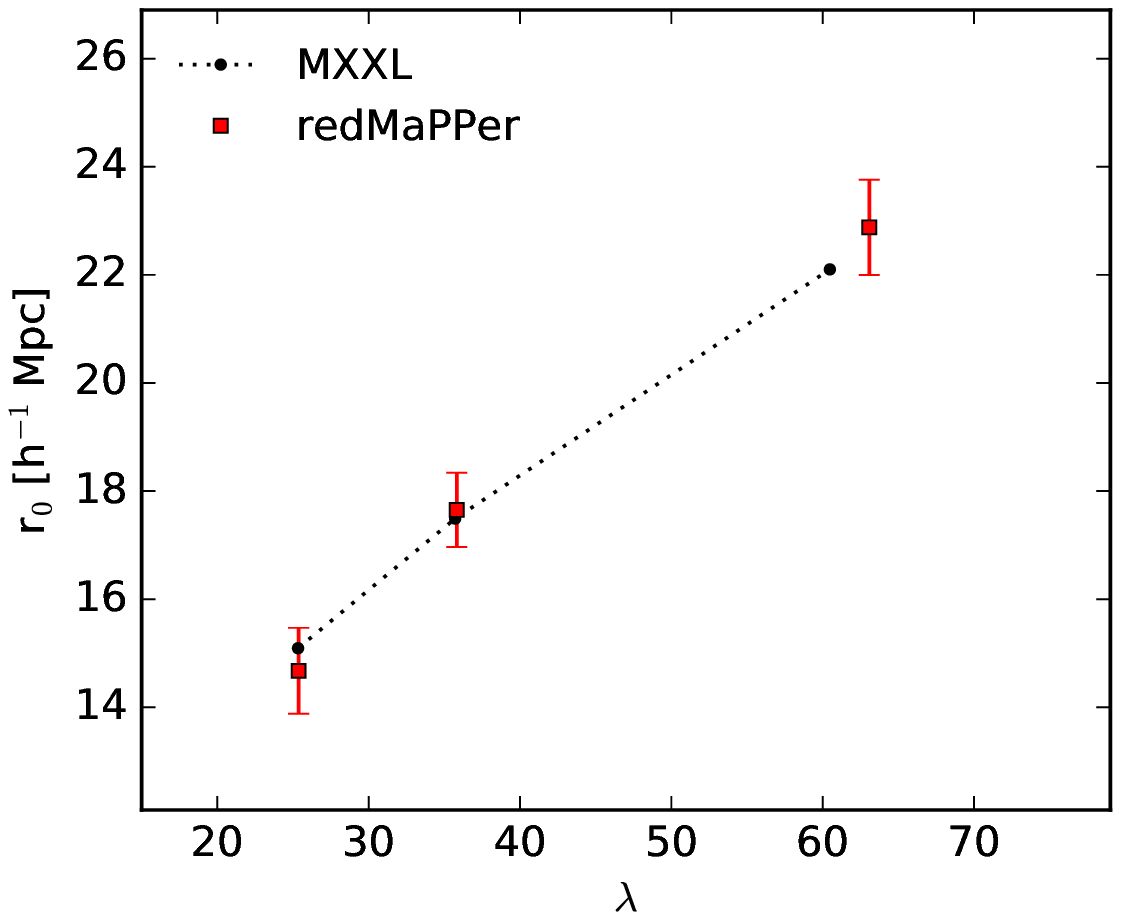}}
\caption{The correlation length $r_0$ obtained from the projected correlation function (red squares) for three redMaPPer richness subsamples, $\lambda \in [22, 30)$, $[30, 45)$, and $[45, 200)$, in the redshift range $0.080\leq z_{\rm clu}\leq0.325$. The dotted black line is the equivalent MXXL correlation length points in the same richness ranges, once the mass--richness relation obtained later in Sec.~\ref{sect:likelihood} is applied, showing consistency between the data and the MXXL simulations of $\lm$.}
\label{fig:Projected_CF_Fit_vs_R}
\end{figure}

Now we make the same measurement for these three richness subsamples for the 
conservative redshift region $0.080\leq z_{\rm clu}\leq0.325$ where the upper redshift limit is
defined by the careful redMaPPer analysis as the limit of their volume complete region for clusters with richness $\lambda > 20$. The results obtained are shown in Fig.~\ref{fig:Projected_CF_Fit_vs_R}, where the relation between correlation length and richness, that we already noticed in Fig.~\ref{fig:1D_CF}, with an obvious rising trend that is very well fitted by MXXL, reflecting the enhanced bias expected for more massive clusters formed in a Gaussian random field \citep{Kaiser1986}.

From these measured values of the correlation length $r_0(\left<\lambda_1\right>)$, $r_0(\left<\lambda_2\right>)$ and $r_0(\left<\lambda_3\right>)$, where $\left<\lambda_1\right>$, $\left<\lambda_2\right>$, and $\left<\lambda_3\right>$ represent the average cluster richness of the three richness subsamples considered, we observe the following linear relation between the value of the richness $\lambda$ and the value of the correlation length $r_0$:
\begin{equation}
 r_0(\lambda) = 9.87 \pm 0.17 + \left( 0.198 \pm 0.004 \right)\lambda\,.
\end{equation}

The data showing such a clear trend with richness rising at the rate that is consistent with the MXXL simulations is very clear evidence for the standard physical understanding of the formation of structure from a Gaussian random field under gravity. 

%%%%%%%%%%%%%%%%%%%%%%%%%%%%%%%%%%%%%%%%%%%%%%%%%%%%%%%%%%%%%%%%%%%%%%%%%%%%%%%
%%%%%%%%%%%%%%%%%%%%%%%%%%%%%%%%%%%%%%%%%%%%%%%%%%%%%%%%%%%%%%%%%%%%%%%%%%%%%%%
\section{Cluster Abundances}
\label{sect:abundances}
%%%%%%%%%%%%%%%%%%%%%%%%%%%%%%%%%%%%%%%%%%%%%%%%%%%%%%%%%%%%%%%%%%%%%%%%%%%%%%%

We proceed now to study the comoving density of clusters as a function of their richness. For this analysis, we limit our sample to the redshift range $0.080\leq z_{\rm clu}\leq0.325$, where the redMaPPer team has established volume completeness. We can also see in Fig.~\ref{fig:redMaPPer_Redshift_Distribution} that this claim is supported by the way the numbers of clusters scale in proportion to the cosmological volume. Above $z=0.35$, the richness calculated by the redMaPPer algorithm is increasingly limited to a diminishing proportion of relatively luminous galaxy members so that an uncertain estimate has to be made to take into account undetected galaxies below the survey magnitude limit, so that richness estimates become more noisy above this redshift; thus, we expect to obtain more robust results working with a cluster redshift cutoff of $z\leq0.325$.

The number of clusters $n_i$ that one may expect to find in the redshift range $[z_{\rm min}, z_{\rm max}]$, and within a richness range $[\lambda_i, \lambda_i+\Delta\lambda]$ is given by:
\begin{equation}
n_i = \Delta\Omega\int^{z_{\rm max}}_{z_{\rm min}}\int^{\lambda_i+\Delta\lambda}_{\lambda_i} dz\,d\lambda\,\frac{dV}{dz}\frac{dN(\lambda)}{dV\,d\lambda}\,,
\end{equation}
where $\Delta\Omega$ is the fraction of the sky covered by the survey, $dV/dz$ is the comoving volume per unit redshift, and $dN/dV/d\lambda$ is the theoretical cluster richness function.

\subsection{Results}

\subsubsection{Comoving densities}

To compute the cluster abundances, we average between several realizations of the redMaPPer richness distribution to take into account the effect on the number of clusters contained in each richness bin that the error on the algorithm richness estimate of each cluster (provided by the redMaPPer catalogue) can introduce. We take the mean values when convergence is obtained, and consider the standard deviation obtained from all these realizations an additional source of systematics, adding it in quadrature to the Poisson noise in each bin.

\begin{figure*}
 \centering
 \includegraphics[width=0.98\textwidth]{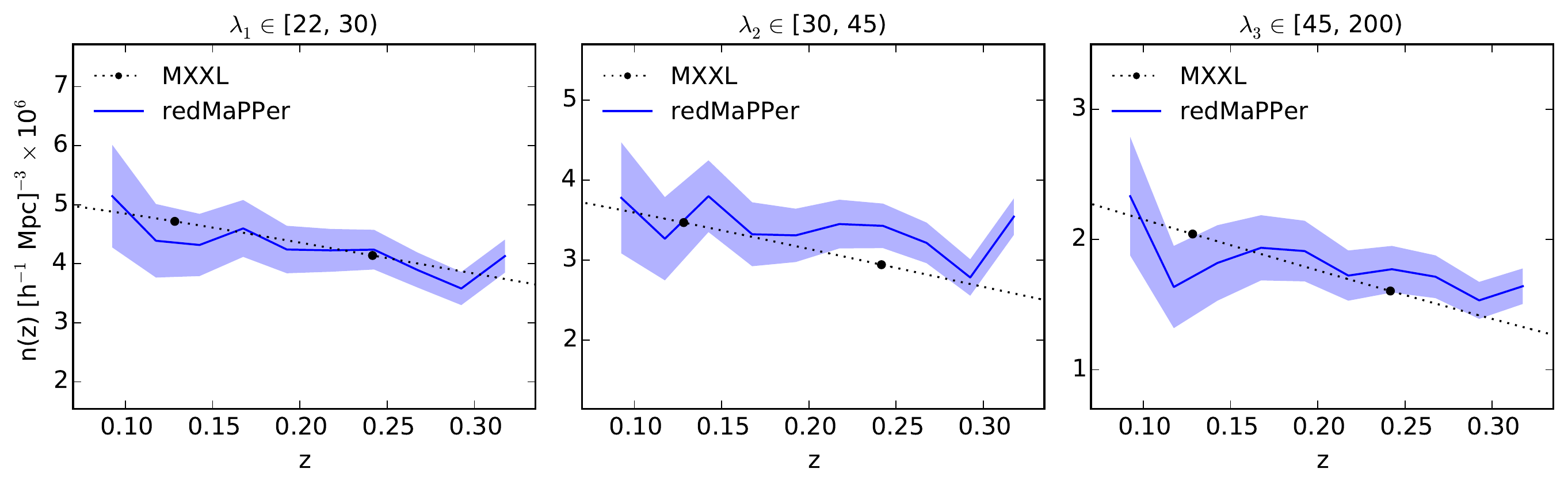}
  \caption{Cluster comoving densities $n(z)$ in the $0.080\leq z_{\rm clu}\leq0.325$ redshift range of three richness subsamples, $\lambda \in [22, 30)$ (left), $[30, 45)$ (middle), and $[45, 200)$ (right). The shaded region represents Poisson noise and errors introduced by the uncertainty in the richness measurement of each cluster, as given by the redMaPPer catalogue. Dotted lines represent the MXXL model density distributions in the same three richness ranges, once the optimal mass--richness relation obtained in Sec.~\ref{sect:likelihood} is used to obtain the synthetic cluster catalogue. A decline is apparent in the data, similar to the predictions of MXXL for this redshift range, corresponding to the expected growth of massive clusters over the past 3 Gyrs.}
 \label{fig:redMaPPer_vs_MXXL_Comoving_Density}
\end{figure*}

The comoving density of clusters $n(z)$ found within the redshift region previously mentioned is shown in Fig.~\ref{fig:redMaPPer_vs_MXXL_Comoving_Density} for three richness samples, where the previously mentioned $\lambda_1\in[22, 30)$, $\lambda_2\in[30, 45)$, and $\lambda_3\in[45, 200)$ binning has been applied, together with the MXXL model abundances.

The agreement between the data and the simulations is within the noise in this redshift range, showing a systematic decline of about 20\%. This is similar to the predictions of MXXL for this redshift range, corresponding to the expected growth of massive clusters over the past 3 Gyrs in the context of $\lm$.

\subsubsection{Richness function}

\begin{figure}
\resizebox{84mm}{!}{\includegraphics{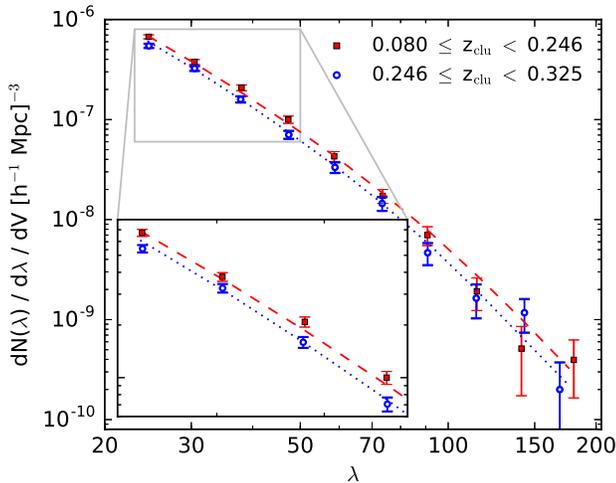}}
\caption{Cluster richness function of the redMaPPer clusters for two redshift subsamples: $0.080\leq z_1<0.246$ with $\left<z_1\right> = 0.186$ (red squares) and $0.246\leq z_2\leq0.325$ with $\left<z_2\right> = 0.287$ (blue circles). These two subsamples are divided into 10 log-spaced richness bins in the range $\lambda \in [22,\,200]$. Error bars include both Poisson noise and the errors on the measurement of the richness of each cluster provided by redMaPPer. A zoomed region $\lambda \in [22, 50]$ shows better the clear evolution with redshift relative to the small errors. The dashed red and dotted blue lines represent the richness functions predicted by the model, that accounts very well for the measured reduction in abundance with redshift. This model is derived in Sec.~\ref{sect:likelihood} from our likelihood analysis based on the MXXL simulations, and the best fit mass--richness relation obtained.
}
\label{fig:redMaPPer_Richness_Function_Evolution}
\end{figure}

To compute the cluster richness function, we restrict ourselves to all the clusters in the redshift range $0.080\leq z_{\rm clu}\leq0.325$, and then we divide them in two redshift bins such that there is approximately equal numbers in each redshift bin. This results in two subsamples with $0.080\leq z_1<0.246$, and $0.246\leq z_2\leq0.325$, with mean redshifts $\left<z_1\right> = 0.186$ and $\left<z_2\right> = 0.287$, respectively. We then consider 10 log-spaced richness bins in the range $\lambda \in [22,\,200]$. The results obtained are shown in Fig.~\ref{fig:redMaPPer_Richness_Function_Evolution}. For comparison we also show the MXXL model results, interpolated to the mean redshifts of the data, $z=0.186$ and $z=0.287$.

It should be noted that, as described in the following section, no evolutionary information was used at all to calibrate the mass--richness relation applied to the MXXL simulations. Thus, the accurate agreement here in Fig.~\ref{fig:redMaPPer_Richness_Function_Evolution} of the data with the evolution predicted by MXXL shows the degree of consistency with the predictions of $\lm$. There was no guarantee that this comparison would reveal the same evolutionary trend in the cluster richness function.

%%%%%%%%%%%%%%%%%%%%%%%%%%%%%%%%%%%%%%%%%%%%%%%%%%%%%%%%%%%%%%%%%%%%%%%%%%%%%%%
%%%%%%%%%%%%%%%%%%%%%%%%%%%%%%%%%%%%%%%%%%%%%%%%%%%%%%%%%%%%%%%%%%%%%%%%%%%%%%%
\section{Likelihood Analysis}
\label{sect:likelihood}
%%%%%%%%%%%%%%%%%%%%%%%%%%%%%%%%%%%%%%%%%%%%%%%%%%%%%%%%%%%%%%%%%%%%%%%%%%%%%%%

We now obtain the mass--richness relation parameter values that best describe the observations through a likelihood analysis. We will be comparing the redMaPPer results with MXXL, covering a wide range of values for a power-law mass--richness relation, where, as we described in Sec.~\ref{sect:mxxl}, $\kappa_{\MR}$ is the pivot mass normalization, $\alpha_{\MR}$ is the slope, and the dispersion is given by $\sigma_{\MR}$. As the assignment of richness given a value of the mass is an stochastic process, we generate several realizations of the synthetic cluster catalogues produced using the MXXL simulations (as we did above in Sec.~\ref{sect:abundances} with the redMaPPer catalogue) averaging the results until convergence is obtained. Note that in this analysis we restrain ourselves to the volume complete region, $0.080\leq z_{\rm clu}\leq0.325$.

For the pivot mass and for the slope we will consider uniform flat priors: $\kappa_{\MR}\,\in\left[1.000,\,2.000 \right]$, and $\alpha_{\MR}\,\in\left[0.500,\,1.600 \right]$. For the case of the scatter, a hard cutoff $\sigma_{\MR}>0$ could bias the resulting posterior distribution, so we adopt the inverse-gamma distribution prior IG$(\epsilon, \epsilon)$ for $\sigma_{\MR}^2$ \citep{Andreon2010, Sereno2015c, Umetsu2016} with $\epsilon$ a very small number (in our case, we take $\epsilon=10^{-3}$). We sample the posterior distribution using a large enough three-dimensional regular fine grid.

\subsection{Clustering}
\label{subsect:clustering_likelihood}

We first obtain an independent mass--richness relation through a likelihood analysis between the clustering data and MXXL. In this case, to obtain the values for $\kappa_{\MR}$, $\alpha_{\MR}$ and $\sigma_{\MR}$ that best describe the clustering results, we rely on the projected correlation function $\Xi(r_\perp)$ of the whole sample, without any redshift nor richness binning. To make the joint analysis more consistent, we limit the sample to the volume complete region $0.080\leq z_{\rm clu}\leq0.325$, as we did in Sec.~\ref{sect:abundances}. We follow the procedure described in Sec.~\ref{sect:correlation} to obtain the points and the covariance matrix required for the likelihood analysis. The likelihood employed has the form:
\begin{equation}
\mathcal{L}(\bm{\Xi}|\bm{\theta},\,\mathcal{C})\propto \exp\left[-\frac{1}{2}\chi^2\left(\bm{\Xi},\,\bm{\theta},\,\mathcal{C}\right)\right]\,,
\end{equation}
where $\bm{\Xi}$ is the 8 dimensional data vector obtained from the cluster sample, $\bm{\theta}$ is the mass--richness relation vector $\bm{\theta}$ = ($\kappa_{\MR}$, $\alpha_{\MR}$, $\sigma_{\MR}$), $\mathcal{C}$ is the $8\times8$ covariance matrix, and
\begin{equation}
\chi^2\left(\bm{\Xi},\,\bm{\theta},\,\mathcal{C}\right) = \left(\bm{\Xi}-\bm{\mu}(\bm{\theta})\right) \mathcal{C}^{-1} \left(\bm{\Xi}-\bm{\mu}(\bm{\theta})\right)^T\,,
\end{equation}
with $\bm{\mu}(\bm{\theta})$ the projected correlation function vector obtained from the MXXL simulations when the mass--richness with $\bm{\theta}$ is considered.

\begin{figure}
\resizebox{84mm}{!}{\includegraphics{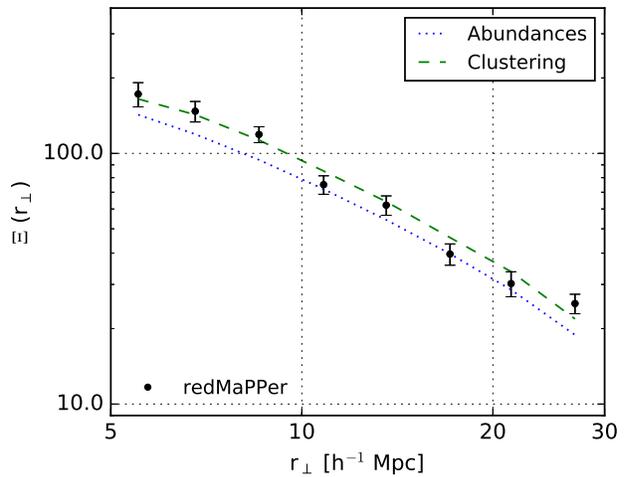}}
\caption{Projected correlation function $\Xi(r_\perp)$ of the redMaPPer clusters for the full $0.080\leq z_{\rm clu}\leq0.325$ redshift range. The dashed green line is the expected correlation function derived for MXXL when we convert between mass and richness using the best fit mass--richness relation from the clustering data alone, clearly there is good self consistency. The blue dotted line represents an independent check using the abundance based mass--richness relation which lies a little below the data at the $2.5\sigma$ level.}
\label{fig:Best_Fit_Correlation_Function}
\end{figure}

The projected correlation function obtained is shown in Fig.~\ref{fig:Best_Fit_Correlation_Function}, together with the MXXL realization with the mass--richness relation parameters given in Table~\ref{tab:Likelihood_Results}.

Finally, we mention that these results are compatible with the ones obtained when we perform the same analysis independently in three richness ranges, the previously mentioned $\lambda_1\in[22, 30)$, $\lambda_2\in[30, 45)$, and $\lambda_3\in[45, 200)$ richness bins. 

\subsection{Abundances}
\label{subsect:abundances_likelihood}

We now obtain the optimal mass--richness relation from the richness function. To compute it, we consider again all the clusters in the redshift range $0.080\leq z_{\rm clu}\leq0.325$, without any redshift binning, and distribute them in 10 log-spaced richness bins in the range $\lambda \in [22,\,200]$ (the maximum richness in this redshift range is found near $\sim200$). The median redshift of the sample considered, which comprises 8,152 clusters, is equal to $\left< z_{\rm clu} \right>=0.24$, and we select the same redshift bin corresponding to the appropriate time step outputted in the MXXL simulations.

To obtain the best fit and the associated confidence intervals, we follow the procedure derived by \cite{Cash1979} for Poisson statistics. To do so, we define the quantity $\mathcal{C}(\bm{n}|\bm{\theta}) = -2\ln \mathcal{L}(\bm{n}|\bm{\theta})$, where $\mathcal{L}(\bm{n}|\bm{\theta})$ is the likelihood function that depends on both the data vector $\bm{n}$, and the mass--richness relation $\bm{\theta}$. The deviations of $\mathcal{C}$ from the minimum follow a $\chi^2$ distribution, and in our case it is equal to:
\begin{equation}
\mathcal{C}(\bm{n}|\bm{\theta}) = -2\ln \mathcal{L}(\bm{n}|\bm{\theta}) = 2 \left(E(\bm{\theta}) - \sum_{i=1}^N n_i \ln e_i(\bm{\theta}) \right)\,,
\end{equation}
where $E(\bm{\theta})$ is the total number of clusters expected in all the $N$ bins once the mass--richness relation with $\bm{\theta}$ = ($\kappa_{\MR}$, $\alpha_{\MR}$, $\sigma_{\MR}$) has been applied to the MXXL simulations, and $n_i$ and $e_i (\bm{\theta})$ are the observed and the expected number of clusters in the bin $i$, respectively.

\begin{figure}
\resizebox{84mm}{!}{\includegraphics{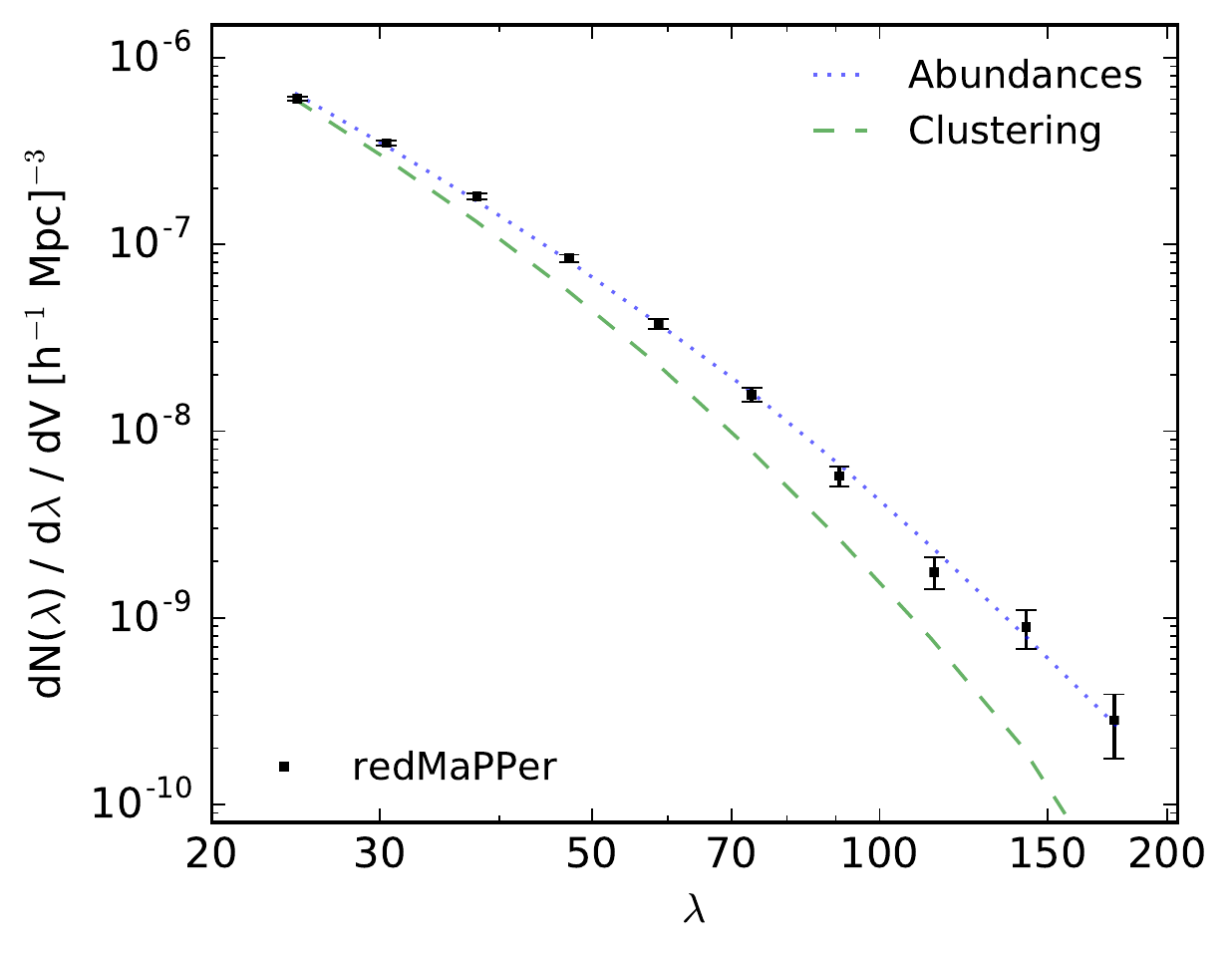}}
\caption{Cluster richness function of the redMaPPer clusters for the full $0.080\leq z_{\rm clu}\leq0.325$ redshift range, obtained dividing the sample in 10 log-spaced richness bins in the range $\lambda \in [22,\,200]$. Error bars combine Poisson errors with the errors on the measurement of the richness of each cluster. The dotted blue and green dashed lines correspond to the richness function of the MXXL realization with the mass--richness relation parameters obtained from the abundances and clustering analysis (Table~\ref{tab:Likelihood_Results}), respectively.}
\label{fig:Best_Fit_Richness_Function}
\end{figure}

The distribution we obtain, together with the MXXL richness function with the mass--richness relation given in Table~\ref{tab:Likelihood_Results}, is shown in Fig.~\ref{fig:Best_Fit_Richness_Function}. This measurement differs from the one described in Sec.~\ref{sect:abundances} and shown in Fig.~\ref{fig:redMaPPer_Richness_Function_Evolution} in that we are only considering one single redshift bin to improve our statistics. A very accurate fit to the data is found for MXXL with the mass--richness relation defined as a power-law, as described and shown in Fig.~\ref{fig:Best_Fit_Richness_Function}. Note that this fit is inherently more accurate than for the correlation function above for which the link between mass and richness is less direct than for abundances as discussed below in Sec.~\ref{sect:conclusions}.

Again, the results obtained through this procedure are compatible with the ones obtained when we perform the same analysis in two redshift subsets of the catalogue with equal number of clusters contained in them, like we did in Sec.~\ref{sect:abundances}, with $0.080\leq z_1<0.246$ and $0.246\leq z_2\leq0.325$.

\subsection{Results and combined analysis}

\begin{figure*}
 \centering
 \includegraphics[width=0.98\textwidth]{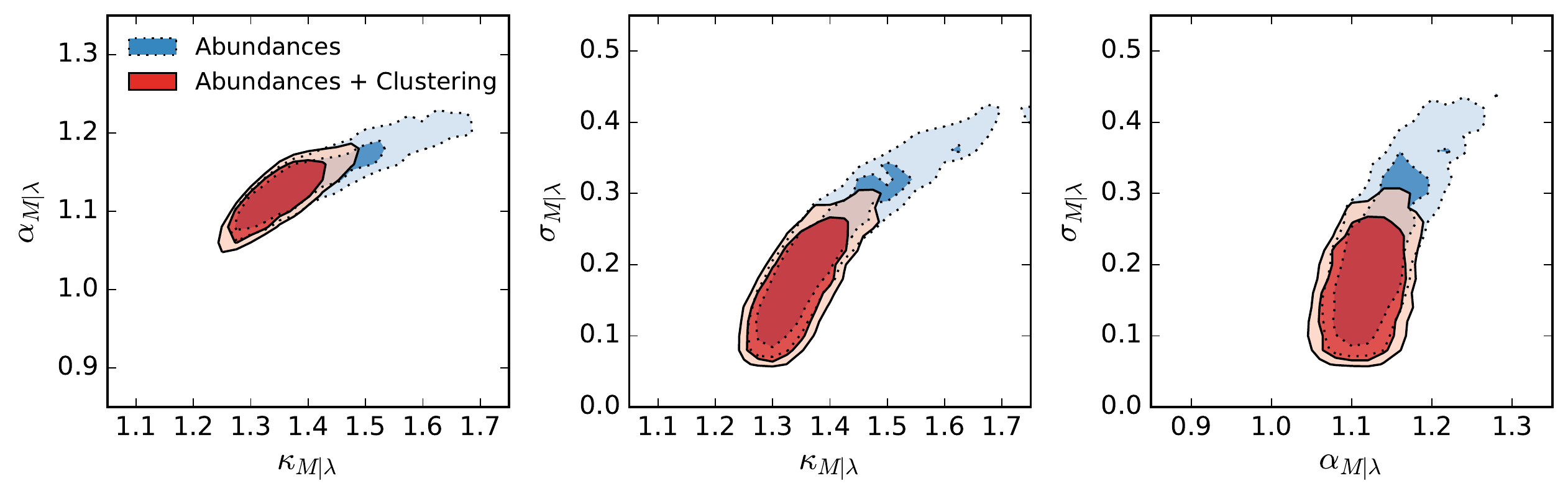}
  \caption{Constraints on the mass--richness relation parameters $\kappa_{\MR}$, $\alpha_{\MR}$ and $\sigma_{\MR}$ at $1\sigma$ and $2\sigma$ confidence levels when the abundances (blue) and the abundances + clustering (red) analysis are performed.}
 \label{fig:Likelihood_Contours}
\end{figure*}

\begin{figure*}
 \centering
 \includegraphics[width=0.98\textwidth]{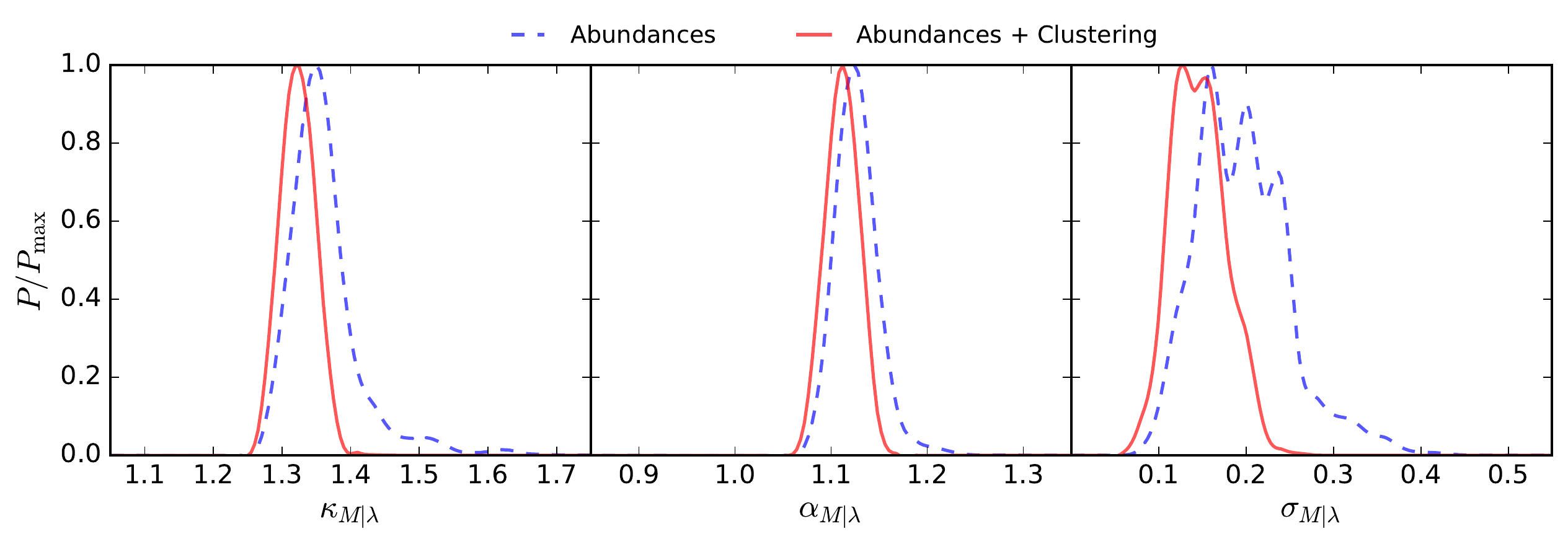}
  \caption{Posterior probability distribution of the mass--richness relation parameters $\kappa_{\MR}$, $\alpha_{\MR}$ and $\sigma_{\MR}$ marginalized over the other 2 parameters when the abundances (dashed blue) data only and the abundances + clustering (solid red) combined data are considered.}
 \label{fig:Likelihood_Marg_Probabilities}
\end{figure*}

The $1\sigma$ and the $2\sigma$ confidence regions of the $\kappa_{\MR}$, $\alpha_{\MR}$ and $\sigma_{\MR}$ values, obtained through the abundance analysis and the joint analysis are shown in Fig.~\ref{fig:Likelihood_Contours}, and the marginalized posterior probabilities of each parameter are given in Fig.~\ref{fig:Likelihood_Marg_Probabilities}. 

To compute the center (mean) and the scale (dispersion) of the marginalized 1D posterior distribution, we use the robust estimators described in \cite{Beers1990}. The results derived from the abundances analysis, the clustering analysis, and the combined analysis are listed in Table~\ref{tab:Likelihood_Results}.

%================================================
\begin{table*}
\center
\caption{
\label{tab:Likelihood_Results}
The confidence values of the marginalized posteriors of the mass--richness relation parameters when the abundances, the clustering, and the combined abundances + clustering analyses are performed.
}
\begin{tabular}{c c c c}
\hline
Parameter	&{\bf Abundances}	&{\bf Clustering}	&{\bf Abundances + Clustering}\\
\hline
\hline
$\kappa_{\MR}$	&$1.351 \pm 0.039$	&$1.548 \pm 0.205$	&$1.341 \pm 0.031$\\	
$\alpha_{\MR}$	&$1.127 \pm 0.021$	&$1.102 \pm 0.197$	&$1.120 \pm 0.017$\\
$\sigma_{\MR}$	&$0.194 \pm 0.053$	&$0.226 \pm 0.089$	&$0.164 \pm 0.039$\\
\hline
\end{tabular}
\end{table*}
%================================================

From our results, it is clear that the abundance analysis is much powerful in means of constraining the mass--richness relation, as the range of parameters that satisfy the clustering observations is much larger in comparison. 

It should be noted, though, that both measurements provide very different mass--richness relations. If we compare how the MXXL synthetic catalogues drawn from both mass--richness relation perform in terms of agreeing with the data, we find that the abundances mass--richness relation (i.e., the one obtained in Sec.~\ref{subsect:abundances_likelihood}) is discrepant with the correlation function results at the $2.4\sigma$ level (Fig.~\ref{fig:Best_Fit_Correlation_Function}). On the other hand, when we apply the clustering mass--richness relation (i.e., the one obtained in Sec.~\ref{subsect:clustering_likelihood}) to the MXXL and measure the richness function, we find that there is a deviation of $8.1\sigma$ with respect to the best fit, making clear that the clustering mass--richness relation is not suitable to describe the cluster abundances found in the data.

In any case, when we combine both analyses and perform a joint likelihood, we obtain a concordance mass--richness relation that describes both measurements, with a pivot mass of:
\begin{equation}
\kappa_{\MR} \equiv \ln \left( \frac{M_{200c}(\lambda_0=60)}{10^{14}\cmass} \right) = 1.341 \pm 0.031\,,
\nonumber
\end{equation}
an slope equal to:
\begin{equation}
\alpha_{\MR} = 1.120 \pm 0.017\,,
\nonumber
\end{equation}
and a scatter:
\begin{equation}
\sigma_{\MR} \equiv \Delta \ln \left( \frac{M_{200c}(\lambda)}{10^{14}\cmass} \right) = 0.164 \pm 0.039\,.
\nonumber
\end{equation}

For comparison with \cite{Simet2016} results, and using \cite{Hu2003} recipe to convert between mass definitions, we can compute the mass obtained through this mass--richness relation at $\lambda=40$:
\begin{equation}
\log_{10} \left(M_{200m}(\lambda=40)\right)=(14.534 \pm 0.015) \cmass\,.
\end{equation}

%%%%%%%%%%%%%%%%%%%%%%%%%%%%%%%%%%%%%%%%%%%%%%%%%%%%%%%%%%%%%%%%%%%%%%%%%%%%%%%
%%%%%%%%%%%%%%%%%%%%%%%%%%%%%%%%%%%%%%%%%%%%%%%%%%%%%%%%%%%%%%%%%%%%%%%%%%%%%%%
\section{Redshift Enhancement}
\label{sect:redshift_enhancement}
%%%%%%%%%%%%%%%%%%%%%%%%%%%%%%%%%%%%%%%%%%%%%%%%%%%%%%%%%%%%%%%%%%%%%%%%%%%%%%%

Finally, as an independent consistency check of the mass--richness relation obtained, we measure the redshift enhancement effect produced by these redMaPPer clusters as a function of their richness, and compare it with the model drawn from such mass--richness relation. This effect is caused by a cluster acting as a gravitational lens with shear $\gamma(r)$ and convergence $\kappa(r)=\Sigma(r)/\Sigma_{\rm crit}$, where $\Sigma(r)$ is the projected mass of the cluster as a function of distance from the center, and $\Sigma_{\rm crit}$ is the critical surface density defined as:
\begin{equation}
 \Sigma_{\rm crit} = \frac{c^2}{4\pi G}\frac{D_s}{D_l\,D_{ls}}\,,
\end{equation}
with $D_s$, $D_l$ and $D_{ls}$ the angular-diameter distances between the observer and the source, the observer and the lens, and the source and the lens, respectively. So, for a lens with a given shear and a convergence, the magnification produced by it is equal to:
\begin{equation}
\mu = \frac{1}{\left(1-\kappa \right)^2-\left|\gamma\right|^2}\,.
\end{equation}
This magnification leads to the so called {\it magnification bias} \citep{Broadhurst1995}, as the number of background lensed sources is modified by a combination of two effects: the luminosity limit of the survey increasing from $L_{\rm lim}$ to $L_{\rm lim}/\mu$ in the lens region, as the fluxes from the sources are amplified, and the number of objects behind the lens decreasing, as sky area behind it is expanded or dilated by a factor $1/\mu$. Thus, the original number of objects observed, $n_0$, is now modified to:
\begin{equation}
 n_{\rm obs}\left[L_{\rm lim}(z)\right]=\frac{1}{\mu}n_0\left[>\frac{L_{\rm lim}(z)}{\mu}\right]\,.
\end{equation}
So, this modification on the number of observed objects, $n_{\rm obs}$, results in an enhancement of the average redshift of these same background objects, given by:
\begin{equation}
 \overline{z}_{\rm back}=\frac{\int n_{\rm obs} (z)\,z\,dz}{\int n_{\rm obs} (z)\,dz}\,.
\end{equation}
So, we define the redshift enhancement $\delta_z$ as:
\begin{equation}\label{eq:redshift_enhancement}
\delta_z(r)\equiv\frac{\overline{z}(r)-\overline{z}_{\rm total}}{\overline{z}_{\rm total}}\,,
\end{equation}
where $\overline{z}(r)$ is the measured average redshift of the lensed $n(r)$ background objects inside a radial bin at a distance $r$ from the center of the lens, and $\overline{z}_{\rm total}$ is the average redshift of the unlensed background objects, as would be seen without a lens. We refer the reader to \cite{Coupon2013} and \cite{Jimeno2015} to a more in-depth description of this measurement. To model this effect, we will also adopt the same assumptions that were considered by them, that is, a NFW profile \cite{Navarro1996}, the mass-concentration relation of \cite{Bhattacharya2013}, and the Schechter parametrization of the V-band luminosity function as given by \cite{Ilbert2005}. Once the redshift enhancement model as a function of cluster mass is obtained, we use the mass--richness relation obtained in Sec.~\ref{sect:likelihood} to convert it into a function of cluster richness.

\subsection{Results}

We proceed now to measure the integrated redshift enhancement for three different richness cluster subsamples $\lambda_1\in[22, 30)$, $\lambda_2\in[30, 45)$, and $\lambda_3\in[45, 200)$ at redshifts below $z<0.400$. To do so, we first measure the average redshift $\overline{z}_{\rm total}$ of all the CMASS galaxies above $z>0.425$, to ensure there is a gap between the clusters and them, and then measure the average redshift of all those background galaxies that lie in the $r_\perp < 0.3 \hdist$ region, where $r_\perp$ is the transverse distance to the CG of the redMaPPer cluster. The redshift enhancement signal $\delta_z$ is then obtained from Eq.~\ref{eq:redshift_enhancement}.

\begin{figure}
\resizebox{84mm}{!}{\includegraphics{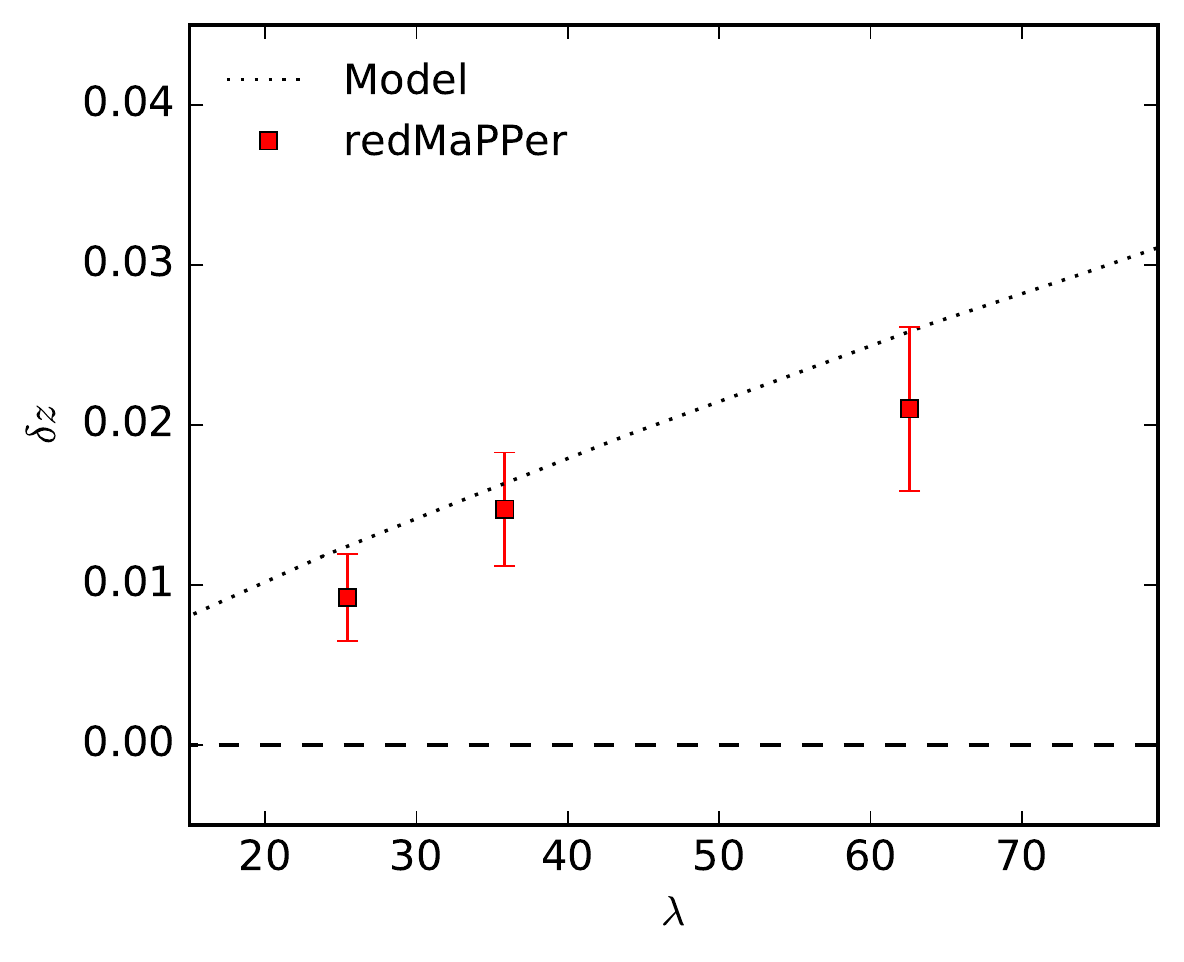}}
\caption{Integrated redshift enhancement signal in the $r_\perp < 0.3 \hdist$ region for three richness redMaPPer subsamples with $\lambda\in[22, 30)$, $[30, 45)$, and $[45, 200)$, in the redshift range $z_{\rm clu}<0.400$. The model prediction as a function of richness is obtained through the mass--richness relation of the joint analysis, as given by Table~\ref{tab:Likelihood_Results}.}
\label{fig:Redshift_Enhancement}
\end{figure}

Finally, for each richness subsample we generate 200 random samples with the same number of clusters, distributed in the same volume and following their same redshift distribution. Then, we measure the integrated redshift enhancement in each of them, and then compute the standard deviation of these 200 measurements to have an estimation of the errors associated to this measurement.

The redshift enhancement signal $\delta_z$ obtained for these three richness subsamples is shown in Fig.~\ref{fig:Redshift_Enhancement}, together with the lensing model whose dependency with richness is obtained from converting mass into richness using the mass--richness relation of the combined analysis, as given in Table~\ref{tab:Likelihood_Results}.

These results are a significant improvement with respect to previous measurements \citep{Coupon2013, Jimeno2015} because of the increase in the numbers of clusters and background galaxies with measured redshifts. The redshift enhancement effect is detected with a significance of $6.6\sigma$ for the whole sample. Also, we observe a clearer trend of higher values of $\delta_z$ for larger values of the average richness of the sample considered, as expected. It should be noted, though, that all the values of $\delta_z$ fall below the model, with a deviation of $\sim1\sigma$ for the richness subsamples $\lambda_1$ and $\lambda_3$. This could be an indication of the level of miscentering present in the position of the CGs with respect to the deepest part of the gravitational potential well, which would be the main cause of the dilution of the redshift enhancement signal.

%%%%%%%%%%%%%%%%%%%%%%%%%%%%%%%%%%%%%%%%%%%%%%%%%%%%%%%%%%%%%%%%%%%%%%%%%%%%%%%
%%%%%%%%%%%%%%%%%%%%%%%%%%%%%%%%%%%%%%%%%%%%%%%%%%%%%%%%%%%%%%%%%%%%%%%%%%%%%%%
\section{Discussion and Conclusions}
\label{sect:conclusions}
%%%%%%%%%%%%%%%%%%%%%%%%%%%%%%%%%%%%%%%%%%%%%%%%%%%%%%%%%%%%%%%%%%%%%%%%%%%%%%%

From the original redMaPPer catalogue and the latest DR12 spectroscopy, we have created a sample of $\sim$ 25,000 clusters with spectroscopic redshifts, and studied the cluster correlation function and the cluster abundances as a function of their optical richness, in the volume complete region $0.080\leqslant z \leqslant 0.350$.

For the correlation function calculation, we have used the centering probabilities of the candidate central galaxies, that the redMaPPer catalogue provides, as an additional weight to obtain a clearer signal. We detect a significant increase of the amplitude of both the monopole and the projected correlation function for higher average richness subsamples, but we do not notice any redshift dependency when those richness subsamples are split into different redshift bins.

On the other hand, when we measure the cluster comoving density on the range $0.080 \leqslant z \leqslant 0.325$, and measure the richness function for two redshift slices, $z_1=0.186$ and $z_2=0.287$, we do detect a clear continuous evolution of the abundances of clusters, which is in excellent agreement with the MXXL simulations.

We have compared our measurements with a synthetic cluster catalogue that we have created from the MXXL simulation, one of the largest cosmological simulations available, that allows us to mimic the distribution, miscentering and peculiar velocities of central galaxies within clusters. To do so, we have assumed the usual power-law mass--richness relation to convert the masses of the DM haloes found in the simulations into richness. Assuming a $\lm$ {\em Planck} cosmology, we find that the best agreement with the data is obtained for a mass--richness relation with a pivot mass $\ln \left( M_{200c}(\lambda_0=60)/10^{14}\cmass \right) = 1.333\pm0.034$, a slope $\alpha_{\MR} = 1.118\pm0.017$, and a scatter $\sigma_{\MR}=0.158\pm0.044$.

\begin{figure}
\resizebox{84mm}{!}{\includegraphics{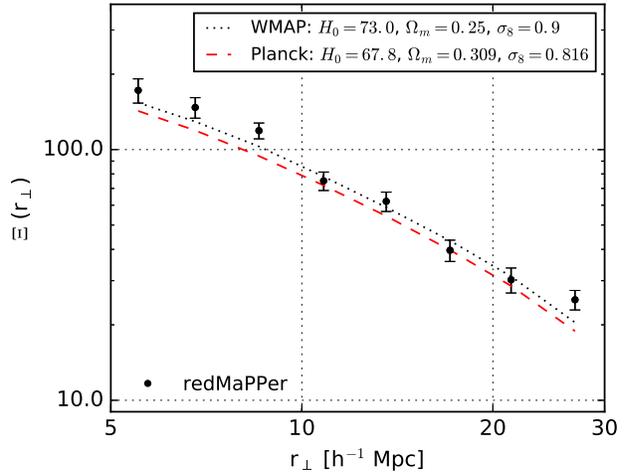}}
\caption{The projected correlation function compared with MXXL derived self consistently for two different cosmologies when the mass--richness obtained from the abundance matching method is used: as the green dotted line, the former WMAP cosmology with $\Omega_m=0.25$ and $\sigma_8=0.9$, and as the blue dashed line, the {\em Planck} cosmology with $\Omega_m=0.309$ and $\sigma_8=0.816$. The WMAP cosmology results are much more in agreement with the observations than the values obtained when the {\em Planck} cosmology is considered.}
\label{fig:Cosmo_Comparison}
\end{figure}

The MXXL fits to the data are remarkably good in all respects as a function of richness, redshift and separation with no clear discrepancy visible in each of these plots, providing very strong confirmation of the detailed viability of the $\lm$ model. However, we do notice that the amplitude of the correlation function is slightly higher than expected when only the abundances are used to constrain the mass--richness relation, with a deviation of $\simeq 2.5\sigma$. The clustering measurements are described by a large range of mass--richness relation parameters, many of them agreeing with the abundance analysis, but the inherent precision is lower for defining the mass--richness relation via clustering. This possible tension between the abundance defined mass--richness relation and the amplitude of the observed correlation function has motivated us to make the same consistency check with the former MXXL and Millennium Simulation WMAP consensus values of $\sigma_8$, $\Omega_m$ and $H_0$ because they were until recently significantly different from the most recent {\em Planck} weighted values and these parameters are the most important for predicting the abundances of clusters and their correlation function. With these former values we find a completely acceptable consistency between the correlation function and the abundances. In Fig.~\ref{fig:Cosmo_Comparison} we show the comparison between the results obtained from both cosmologies when we use the mass--richness relation obtained from the abundance matching technique in both cosmologies. This may prove very interesting because of the increased support for $H_0=73$ km s$^{-1}$ from the independent local distance ladder measurements \citep{Riess2016} that adds support to previous claims for this value at somewhat lower significance \citep{Freedman2001} and from independent lensing time delay estimates of $H_0$ \citep{Bonvin2016}.

Finally, we measured the redshift enhancement effect produced by the gravitational magnification of the redMaPPer clusters on the background CMASS subsample of BOSS galaxies. We found once again a clear relation between the richness and the amplitude of the gravitational lens magnification effect, with consistency at the $1.1\sigma$ level, a little below the model prediction, which could be produced by the intrinsic miscentering of central galaxies with respect to the center of the projected gravitational potential well and will be examined more thoroughly in the future. 

\begin{figure}
\resizebox{84mm}{!}{\includegraphics{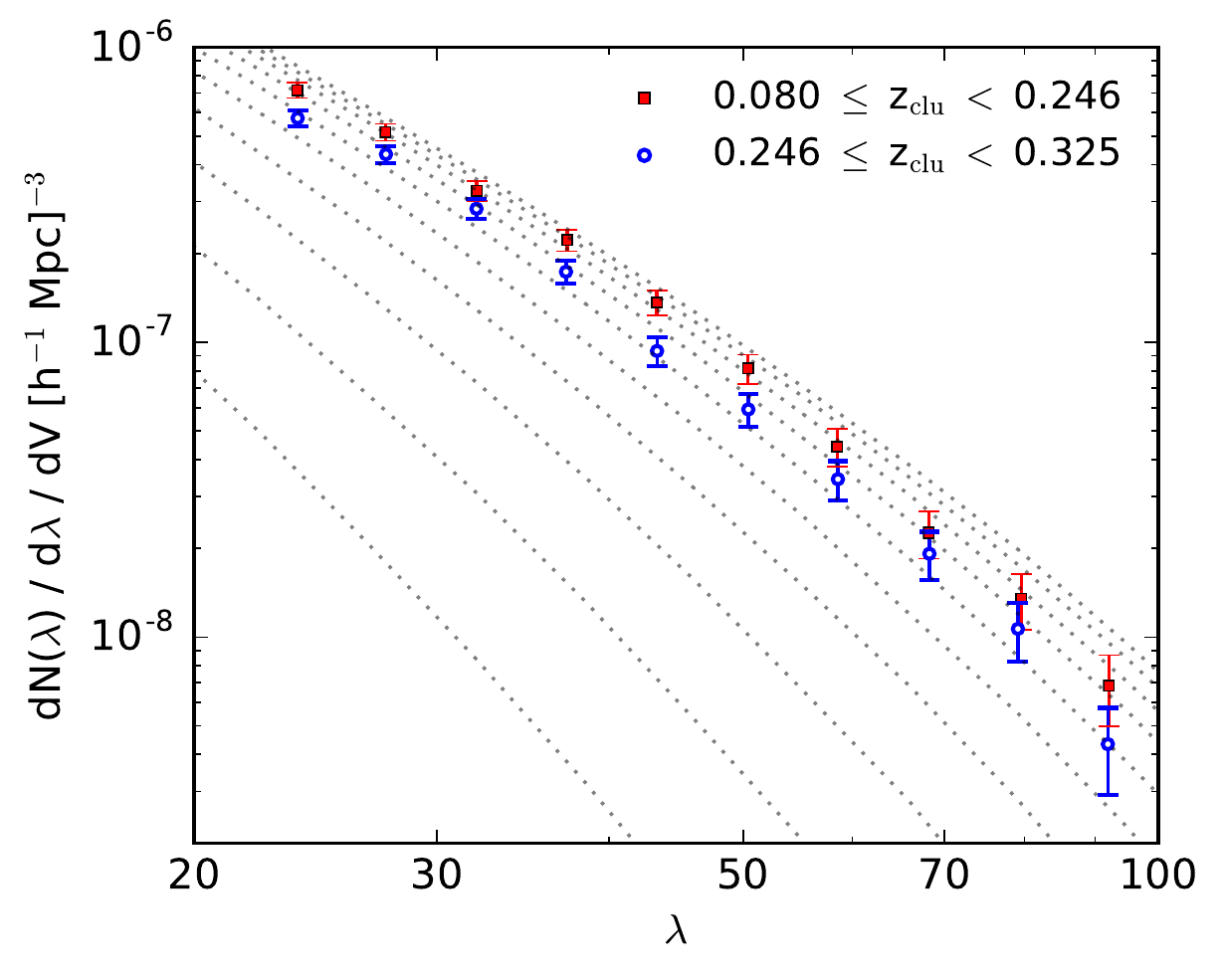}}
\caption{The model curves show the MXXL richness function for snapshots with redshift (from top to bottom): 0.027, 0.083, 0.154, 0.242, 0.351, 0.486, 0.652, 0.857, 1.110, 1.424. The mass--richness relation of the joint analysis, as given by Table~\ref{tab:Likelihood_Results}, has been used to convert between mass and richness. For comparison, the cluster richness function of the redMaPPer clusters in the range $22 < \lambda < 100$ and for two redshift subsamples are shown: $0.080\leq z_1<0.246$ with $\left<z_1\right> = 0.186$ (red squares) and $0.246\leq z_2\leq0.325$ with $\left<z_2\right> = 0.287$ (blue circles), illustrating the enormous potential of the upcoming deeper imaging surveys for evaluating the growth of structure.}
\label{fig:MXXL_Richness_Function_Evolution}
\end{figure}

The J-PAS survey of the North now coming on line \citep{Benitez2014} will provide lensing masses and redshifts for all SDSS clusters and will go beyond in redshift, measuring unprecedentedly accurate growth as a function of cluster mass in the range $z \lesssim 1.0$ (Fig.~\ref{fig:MXXL_Richness_Function_Evolution}), with the prospect of constraining the total relativistic species contribution to $\Omega_m$, and obtaining a clearer insight into the emerging tensions between the parameters describing the standard $\lm$ model.

%%%%%%%%%%%%%%%%%%%%%%%%%%%%%%%%%%%%%%%%%%%%%%%%%%%%%%%%%%%%%%%%%%%%%%%%%%%%%%%
%%%%%%%%%%%%%%%%%%%%%%%%%%%%%%%%%%%%%%%%%%%%%%%%%%%%%%%%%%%%%%%%%%%%%%%%%%%%%%%
\hspace{0pt}\\ \textbf{Acknowledgements}\\

TJB is supported by IKERBASQUE, the Basque Foundation for Science.
RL is supported by the Spanish Ministry of Economy and Competitiveness through research projects FIS2010-15492 and Consolider EPI CSD2010-00064, and the University of the Basque Country UPV/EHU under program UFI 11/55.
PJ acknowledges financial support from the Basque Government grant BFI-2012-349.
TJB, RL and PJ are also supported by the Basque Government grant for the GIC IT956-16 research group.
REA acknowledges support from AYA2015-66211-C2-2.
J.M.D acknowledges support of the projects AYA2015-64508-P (MINECO/FEDER, UE), AYA2012-39475-C02-01 and the consolider project CSD2010-00064 funded by the Ministerio de Economia y Competitividad.
KU acknowledges partial support from the Ministry of Science and Technology of Taiwan (grants MOST 103-2112-M-001-030-MY3 and MOST 103-2112-M-001-003-MY3).

Funding for SDSS-III has been provided by the Alfred P. Sloan Foundation, the Participating Institutions, the National Science Foundation, and the U.S. Department of Energy Office of Science. The SDSS-III web site is http://www.sdss3.org/. SDSS-III is managed by the Astrophysical Research Consortium for the Participating Institutions of the SDSS-III Collaboration including the
University of Arizona,
the Brazilian Participation Group,
Brookhaven National Laboratory,
University of Cambridge,
Carnegie Mellon University,
University of Florida,
the French Participation Group,
the German Participation Group,
Harvard University,
the Instituto de Astrofisica de Canarias,
the Michigan State/Notre Dame/JINA Participation Group,
Johns Hopkins University,
Lawrence Berkeley National Laboratory,
Max Planck Institute for Astrophysics,
Max Planck Institute for Extraterrestrial Physics,
New Mexico State University,
New York University,
Ohio State University,
Pennsylvania State University,
University of Portsmouth,
Princeton University,
the Spanish Participation Group,
University of Tokyo,
University of Utah,
Vanderbilt University,
University of Virginia,
University of Washington,
and Yale University.

\bibliography{BibRef}

\label{lastpage}

\end{document}